\documentclass[a4paper,12pt]{article}

\usepackage[a4paper]{geometry}
\usepackage[dvipsnames]{xcolor}
\usepackage[utf8]{inputenc} 
\usepackage[hidelinks]{hyperref}
\usepackage{float}
\usepackage{amssymb,amsmath,amsthm}
\usepackage{amsfonts}

\usepackage[draft,colorinlistoftodos,shadow, textwidth=2.5cm, textsize=scriptsize]{todonotes}
\pdfoutput=1
\usepackage{graphicx}
\usepackage{graphicx, float, tabularx, booktabs}
\usepackage{color}
\usepackage{enumerate}
\usepackage[american]{babel}
\usepackage{stackrel}

\usepackage[numbers, compress]{natbib}
\usepackage{natbib}
\newcommand{\beq}{\begin{equation}}
\newcommand{\eeq}{\end{equation}}
\newcommand{\bea}{\begin{eqnarray}}
\newcommand{\eea}{\end{eqnarray}}
\newcommand{\be}{\begin{equation}}
\newcommand{\ee}{\end{equation}}
\newcommand{\ba}{\begin{eqnarray}}
\newcommand{\ea}{\end{eqnarray}}

\newcommand{\Mpl}{M_{\rm P}}

\newcommand{\ellp}{\ell_{\rm P}}

\newcommand{\mpl}{M_{\rm P}}

\newcommand{\gn}{G}

\usepackage{slashed}

\newcommand{\RC}{r_{\rm C}}
\newcommand{\RS}{r_{\rm S}}

\usepackage{amsfonts}
\usepackage{epsfig}

\usepackage{graphicx, tabularx}
\vspace{-0.1cm}
\renewcommand{\thefootnote}{\fnsymbol{footnote}}

\begin{document}

\begin{center}
{\Large \textbf{
Reconciling microscopic and macroscopic tests of the 
Compton-Schwarzschild correspondence
}}
\end{center}

\vspace{-0.1cm}

\begin{center}
Bernard Carr, $^{a}$\footnote{%
E-mail: \texttt{b.j.carr@qmul.ac.uk} },  Jonas Mureika$^{b,c}$\footnote{%
E-mail: \texttt{jmureika@lmu.edu} } and Piero Nicolini $^{d,e,f}$\footnote{%
E-mail: \texttt{piero.nicolini@units.it} }

\vspace{.6truecm}

\emph{\small  $^a$School of Physics and Astronomy, %\\[0pt]
 Queen Mary University of London,\\[-0.5ex]
\small  Mile End Road, London E1 4NS, UK}\\[1ex]

\emph{\small  $^b$Department of Physics, Loyola Marymount University,\\[-0.5ex] Los Angeles,  California, USA}\\[1ex]

\emph{\small  $^c$Kavli Institute for Theoretical Physics,\\[-0.5ex]
\small University of California Santa Barbara, California, USA}\\[1ex]

\emph{\small  $^d$Dipartimento di Fisica, Università degli Studi di Trieste, \\[-0.5ex]
and Istituto Nazionale di Fisica Nucleare (INFN), Sezione di Trieste, Trieste, Italy
}\\[1ex]

\emph{\small  $^e$Institut f\"ur Theoretische Physik,
Johann Wolfgang Goethe-Universit\"at, \\[-0.5ex] 
and Frankfurt Institute for Advanced Studies (FIAS),
Frankfurt am Main, Germany}\\[1ex]

\emph{\small  $^f$Center for Astro, Particle and Planetary Physics,  New York University Abu
Dhabi, 
\\[-0.5ex]Abu Dhabi, 
United Arab Emirates}\\[1ex]

\end{center}

\begin{abstract}
\noindent{\small } \noindent
We review the experimental constraints on the parameter $\alpha$ associated with the Generalized Uncertainty Principle (GUP) and the parameter $\beta$ associated with the Generalized Event Horizon (GEH).  The Compton-Schwarzschild correspondence implies a relationship between $\alpha$ and $\beta$, with both parameters being of order 1. This presents a problem for our previous `$M+1/M$' model since the extra gravitational force at sub-Planckian masses contravenes observations.  Various resolutions of this problem are discussed.
\end{abstract}
%\pacs{04.70.Dy, 04.60.-m, 04.60.Kz} 	
%\maketitle

\renewcommand{\thefootnote}{\arabic{footnote}} \setcounter{footnote}{0}
%\thispagestyle{empty}
%\clearpage
%\tableofcontents

\section{Introduction}
In previous papers we have explored the possible link between the Generalised Uncertainty Principle (GUP), which 
involves sub-Planckian energies and modifies the expression for the
Compton wavelength, and the Generalized Event Horizon (GEH), which 
involves super-Planckian energies and modifies the expression for the Schwarzschild radius. This gives rise to the Black Hole Uncertainty Principle (BHUP) or Compton-Schwarzschild (CS) correspondence, which 
leads to a unified expression for the generalised Compton and Schwarzschild expressions.  In particular,  we proposed what we term a `$M+1/M$' model which has the appropriate asymptotic behaviour  in the large and small $M$ limit.
If correct, this implies a fundamental link between  sub-Planckian and super-Planckian  
behaviour, with quantum phenomena being interpreted in terms of relativity theory (eg. sub-Planckian black holes) and  gravitational phenomena being interpreted in terms of quantum theory (eg.  Hawking radiation). This would clearly be a crucial feature  of any final theory of quantum gravity. 

The GUP applies in the quantum regime and  involves a parameter $\alpha$, while the  GEH applies in the gravitational regime and involves a parameter $\beta$ \footnote{Not all papers use this notation and some authors denote the GUP parameter by $\beta$.}.  Constraints on $\alpha$ involve accelerator and laser-interferometer experiments which probe the microphysical domain,  whereas constraints on $\beta$ involve measurements of the gravitational force and  possible deviations from the Coulomb law which probe the macrophysical domain. 
In this paper we review these constraints and show that they generally imply an upper limit on $\alpha$ well above 1
and a lower limit on $\beta$ well below 1. In principle, these constraints could be independent but the BHUP or CS correspondence implies that $\alpha$ and $\beta$ are related,  which links the different types  experimental tests. In particular,  in the `$M+1/M$' model,  we argue that one expects $\alpha \sim 1/ \beta$. This implies $\alpha \sim \beta \sim 1$, which is consistent with the current limits but still below detectability.  

In its simplest form, the `$M+1/M$' model implies an extra gravitational force for sub-Planckian masses 
which is inconsistent with observation.  For example,  it corresponds to a force between protons which is larger than the usual gravitational force by a factor of $(\Mpl/m_{\rm p})^2 \sim 10^{40}$ and possibly $(\Mpl/m_{\rm p})^4 \sim 10^{80}$, which would disturb measurements of the other interactions between them. This poses a challenge for  the model since - while there is no direct experimental evidence that Newtonian gravity operates between elementary particles at all - the whole purpose of our proposal is to link gravitational and quantum interactions. Therefore
another aim of this paper is consider possible resolutions of this problem. 

We consider various possibilities.
The first is that the extra gravitational force {\it becomes} the other forces, this involving a rather radical revision of the GUP in which the parameter $\alpha$ depends on the nature of the particle.  A second possibility is that the meaning of the mass $M$ must be revised
and this requires consideration of what are termed the `ADM' and `irreducible' mass of a black hole.  The third possibility is that the `$M+1/M$' model involves two asympotically flat spaces which are connected by a wormhole, with the physical radius $R$ being related to the coordinate radius $r$ by $R \sim r + 1/r$. This avoids the singularity at $r=0$ and means that the gravitational force associated with the $1/M$ term is in some sense pushed into the other space, with the black hole masses being different in the two spaces.
A fourth possibility is that the extra Newtonian force is modified below some scale, either by imposing an exponential cut-off at small masses or 
 by introducing higher-dimensional effects. 
All these models entail some link between quantum and gravitational physics, with elementary particles being regarded as sub-Planckian black holes (cf.  the `strong gravity' model of the 1970s).  We discuss this proposal further in a separate paper.

The paper is organized as follows.  In Sec.~\ref{review1} we review the argument for the Compton-Schwarzschild correspondence, with particular emphasis on our `$M+1/M$' model.  In Sec.~\ref{sec:wfl}
 we present bounds on relevant parameters from recent experiments for super-Planckian and sub-Planckian  masses. 
In Sec.~\ref{sec5} we discuss the problems which arise when one combines these limits and consider possible modifications of the scenario required to overcome these.
 In Sec.~\ref{concl} we draw some conclusions.

\section{Review of `M+1/M' model}
\label{review1}

General Relativity (GR) is a triumphant classical theory of gravitation that has been observationally verified for over a century. Still, there are features of the formulation that must be corrected. Most importantly, GR predicts the existence of a singularity at the centre of a black hole where the spacetime curvature becomes infinite. Since this is the regime where quantum gravity is believed to dominate, it is generally believed that the singularity will be replaced by some other quantum structure. Exactly how this manifests itself, however, is unknown. One such common feature of many candidate quantum gravity theories is the emergence of a minimal length, below which it is not possible to probe physics and characteristics of spacetime \cite{Gar95,Pad97}.  This arises in the context of 
non-commutative geometry 
\cite{Nic09} and it may also feature in models involving gravitational self-completeness \cite{DFG11} and loop quantum gravity \cite{Rov98}.
 A comprehensive overview of minimal length models can be found in Refs.~\cite{SNB12,Hos13}. 

The notion of a minimal length also emerges 
from the Generalized Uncertainty Principle (GUP) \cite{ACV87,ACV88,ACV89,Mag93,KMM95,AdS99,ACS01,ChA03,Adl10,IMN13,CMP11, Carr16, CMN15,CMMN20}, which can be thought of as an extension of the standard Heisenberg Uncertainty Principle (HUP) with a gravitational uncertainty,
\beq
\Delta x\Delta p\geq \frac{\hbar}{2} +\frac{\alpha \gn\Delta p^2}{c^3}
\Rightarrow
 \Delta x \geq \frac{\hbar}{2}\left[ \frac{1}{\Delta p} + \frac{2 {\alpha} \Delta p}{\Mpl^2 c^2}\right] \, .
\label{eq:gup}
\eeq 
where $\Mpl =\sqrt{\hbar c/G} \sim 10^{-5}$g is the Planck mass.
The dimensionless constant $\alpha$ represents the number of Planck lengths (squared) by which the gravitational contribution alters the HUP uncertainty.  This is because 
the associated minimal length scale 
is $(\Delta x)_\mathrm{min} \sim \alpha^{1/2} \ellp$, where  $\ellp = \sqrt{\hbar G/c^3} \sim 10^{-33}$cm is the Planck length.

An ideal theory of quantum gravity should be able to seamlessly meld the microscopic regime of quantum mechanics with the macroscopic domain of General Relativity. Each can be described by their respective characteristic length scales, 
which are the Compton wavelength $\RC = \hbar/(Mc)$ and the Schwarzschild radius $\RS = 2GM/c^2$. The former is associated with particles and the latter with black holes. They converge when
\beq
r_{\rm S} = r_{\rm C} ~~~\Longrightarrow~~~
 M_{\rm min} = \Mpl/\sqrt{2} , ~~~
r_{\rm min} = \sqrt{2} \, \ellp \,.
\label{SC}
\eeq
This may be interpreted as gravitational ``self-completeness'': any attempt to probe a particle above the Planck energy will result in the formation of a black hole, so that one is actually probing the Schwarzschild radius instead. 
The self-completeness paradigm suggests that the minimum $M_{\rm min}$ corresponds to 
either the heaviest particle if approached from the left or the smallest black hole if approached from the right. 
In this sense,  it is a critical point or a phase transition between 
 black holes and particles.  

Of course, the expressions for $\RC$ and $\RS$ - and hence Eq.~\eqref{SC} - may not be valid down to the Planck scale itself.
In particular, Adler has suggested \cite{AdS99} that, as the minimum is approached from the left (particle side), the Heisenberg Uncertainty Principle should be replaced by the GUP.  It was argued in Ref.~\cite{Carr16} that this 
leads to a modified  
Compton wavelength\footnote{As discussed later, this expression omits a factor of $2$.}
\beq
\RC' 
= \frac{\hbar}{Mc} \left[ 1  + \alpha  \left( \frac{M}{\Mpl} \right)^2 \right] ~~~ (M < \Mpl) \, ,
\label{RC'}
\eeq 
provided the substitutions $\Delta x = \RC'$ and $\Delta p= Mc$ are made.
Conversely, it has been argued that approaching the minimum from the right (black hole side)
requires the Schwarzschild radius to be replaced by a generalized event horizon (GEH),
\beq
 \RS' = \left( \frac{2 GM}{c^2} \right) \left[ 1  + \frac{\beta}{2} \left( \frac{\Mpl}{M} \right)^2 \right] ~~~ (M > \Mpl)\, ,
 \label{RS'}
\eeq
 where $\beta$ is another dimensionless constant. 
 Setting $\RS' = \RC'$, one finds a minimum mass of
\beq
M_{\rm min} = 
 \sqrt{\frac{\beta-1}{\alpha - 2}} \, M_{\rm Pl} \,, \quad r_{\rm min} = \frac{2 - \alpha  \beta}{\sqrt{(\alpha - 2)(\beta -1)}} \, \ellp \, , 
\label{Planck}
\eeq
which reduces to Eq.~(\ref{SC}) when $\alpha =  \beta = 0$. 
However, it is possible that Eqs.~(\ref{RC'}) and (\ref{RS'}) 
merely give the lowest order terms of some higher theory, which would necessitate a modification of Eq.~\eqref{Planck}.

The similarity of Eqs.~(\ref{RC'}) and (\ref{RS'}) hints that there is a 
fundamental link between particles on small scales and black holes on large scales, where the discontinuous phase transition is replaced by a smooth minimum. This is dubbed the Black Hole Uncertainty Principle (BHUP) correspondence \cite{CMN15} or
Compton-Schwarzschild (CS) correspondence  \cite{Lake:2015pma} and leads to
a unified expression for the sub-Planckian ($M<\Mpl$) and super-Planckian ($M>\Mpl$) mass scales.
The simplest example of this arises if $\alpha =2$ and $\beta =1$, since Eqs.~\eqref{RC'} and \eqref{RS'} 
are then  identical.  Such a parameter-free model might seem implausible but is not excluded experimentally.   

However, the status of Eq.~\eqref{RC'} needs to be examined more carefully because the identifications $\Delta x = \RC'$ and $\Delta p= Mc$ should only be regarded as order-of-magnitude relationships.  Indeed,  if they were precise, Eq.~\eqref{eq:gup} would imply
\beq
\RC' 
= \frac{\hbar}{2Mc} \left[ 1  + 2 \alpha  \left( \frac{M}{\Mpl} \right)^2 \right] ~~~ (M < \Mpl) \, 
\label{RC''}
\eeq 
and Eq.~\eqref{Planck} would then become
\beq
M_{\rm min} = 
 \sqrt{\frac{\beta-1}{2\alpha - 2}} \, \Mpl \, \quad r_{\rm min} = \frac{ \sqrt{2} \, (1 - \alpha  \beta)}{\sqrt{(\alpha - 1)(\beta -1)}} \, \ellp \, .
\label{Planck'}
\eeq
In this case,  Eq.~\eqref{RC''} no longer gives the standard Compton expression and it is only consistent with the Schwarzschild expression in the super-Planckian regime for $\alpha =2$.
Although the constant in the GUP itself is exact, 
the issue is how one ``translates'' from $\Delta x$ and $\Delta p$ to $\RS$ and $M$. 

For a more precise analysis, let us assume $\Delta x = \gamma r_{\rm C}'$ and $\Delta p = \epsilon M$ for some constants $\gamma$ and $\epsilon$.  Putting $c= \hbar = 1$, Eq.~\eqref{eq:gup} implies that the expression for the Compton wavelength becomes
\be
 r_{\rm C}' = \frac{1}{2 \gamma \epsilon M} + \frac{\alpha \epsilon}{\gamma} GM \, .
\ee
Therefore $r_{\rm C}' \equiv r_{\rm S}'$ 
 only if
\be 
 \alpha \epsilon = 2 \gamma,  \quad  \beta = 1/(2 \gamma \epsilon) = 1/(\alpha \epsilon^2) \,.  
\label{alphabeta}
\ee
Since the Compton wavelength itself arises in multiple contexts, it is hard to specify $\gamma$ precisely.  However,  it might seem reasonable to put $\epsilon = 1$, in which case $\gamma = \alpha/2$ (so the GUP parameter relates to the uncertainty in position) and $\beta = 1/\alpha$.  

These considerations suggest that the most natural expression for the unified Compon-Schwarzschild scales is
\beq
r_{\rm CS} = \frac{\beta \hbar}{Mc} + \frac{2GM}{c^2} \, ,
\label{unified'}
\eeq
which has a smooth minimum for $\beta > 0$ and is equivalent to Eq.~(\ref{RS'}) but
holds for both $M < \Mpl$ and $M > \Mpl$. 
In contrast to Eq.~(\ref{RC'}), the free parameter is now associated with the first term instead of the second. 
It has been argued that this is
plausible because the constant in the expression for the Compton scale depends on the physical context~\cite{LaC18} \footnote{The Compton wavelength
first appeared historically in the expression for the Compton cross-section in the scattering of photons off electrons.
Subsequently, it has arisen in various other contexts.  For example, it is relevant to processes which involve turning photon energy 
 into rest mass energy 
and the {\it reduced} Compton wavelength 
 appears naturally in the Klein-Gordon and Dirac equations. 
One can also associate it
with the {\it localisation} of a particle, there being 
both non-relativistic and relativistic arguments for this notion.}.  If one imposed $\beta =1$, so that the usual Compton expression applies,  Eq.~\eqref{alphabeta} would require $\alpha = 1/ \epsilon^2$.  However, associating the free parameter with the second term is less natural since the expression for the Schwarzschild radius is exact. 
We will later argue that both $\alpha$ and $\beta$ are $O(1)$, so one is necessarily close to the simplest case with $\alpha =1$ and $\beta =1$. 

\begin{figure}[h]
\begin{center}
\includegraphics[scale=.2]
{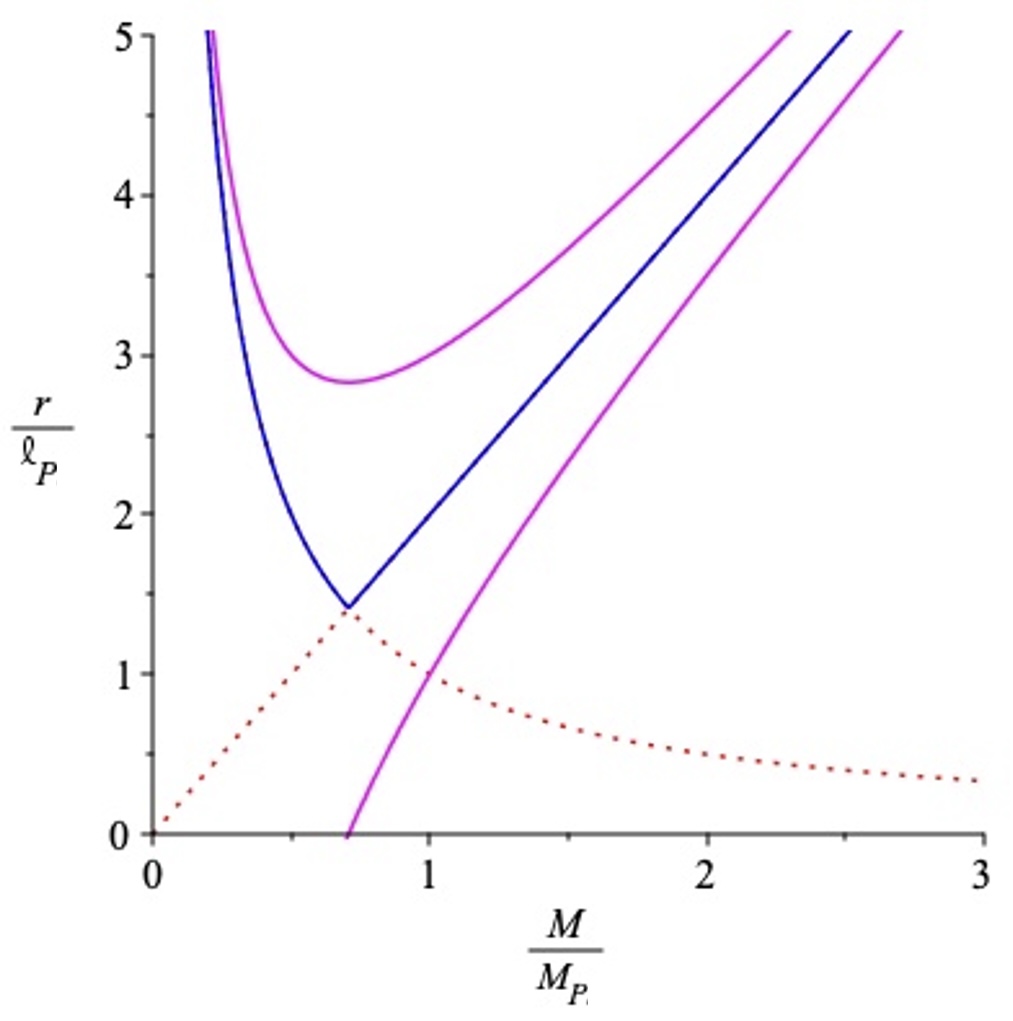}
\caption{
The blue lines show 
the Schwarzschild and Compton scales, their 
 intersection giving 
the smallest resolvable length scale.
The red dotted lines give the 
(inaccessible) continuations of these
 curves. The upper 
magenta curve shows the unified
Compton and Schwarzschild scale ($r_{\rm CS}$) for $\beta > 0$.
The lower 
magenta curve applies for $\beta < 0$ but this does not provide a unification.}
 \label{SCfig}  
 \end{center}     
\end{figure}  

 We previously suggested a simple realization of this proposal  \cite{CMN15}, in which the bare mass $M$ and ADM mass $M_{\rm ADM}$ of a system are related by
\beq
M_{\rm ADM} 
= M +  \frac{\beta \Mpl^2}{2M} \,.
\label{madm}
\eeq
In the super-Planckian regime ($M \gg \Mpl$) one has $M_{\rm ADM} \approx M$, but in the sub-Planckian limit ($M \ll \Mpl$) it scales as $1/M$. 
 Because of this behaviour, we dub this the `$M+1/M$' model.  For $\beta > 0$, $M_{\rm ADM}$ has a minimum value of 
 $ \sqrt{2 \beta} \, \Mpl$ at $M = \sqrt{\beta/2} \, \Mpl$.  Note that Eq.~\eqref{madm} shares some similarity with the `quantum-N portrait' approach of Dvali {\it et al.} \cite{DvG12, DvG13,DvG13+} with the $1/M$ term in the super-Planckian regime being associated with some form of quantum mechanical hair~\cite{NiS14,Nicolini:2023hub}. 
As various authors \cite{tGR+18} (including 
Giddings \cite{Gid17}) have claimed, the latter corresponds to a quantum atmosphere visible at scales conventionally within  reach of modern telescopes.

The ADM mass determines the spacetime structure around the mass $M$ 
by modifying the Schwarzschild metric:
\bea
g_{00} = 1-\frac{2M_{\rm ADM}}{\Mpl^2r} \,. 
\label{newmetric}
\eea
The corresponding horizon radius
is then
\bea
r_{\rm CS} = \frac{2M}{\Mpl^2} +\frac{\beta}{M}
\approx
\begin{cases}
\frac{2M}{\Mpl^2} & (M \gg \Mpl) \label{bigr}\\
 \frac{2+\beta}{\Mpl} & (M \approx \Mpl) \label{ellpr}\\
\frac{\beta}{M} & (M \ll \Mpl ) .  \label{smallr}
\end{cases}
\label{horizon}
\eea
where we have put $\hbar = c = 1$. This is in agreement with Eq.~\eqref{unified'} and
 illustrated by the upper magenta curve in Fig.~\ref{SCfig}.  For $\beta < 0$, the curve terminates at $M = \sqrt{|\beta|/2} \, \Mpl$ (lower magenta curve in Fig.~\ref{SCfig}).  At this point 
$r_{\rm CS} = 0$, which suggests a connection with
models involving asymptotic safety \cite{BoR06}.  However,  there is no corresponding Compton curve in this case,  so we will restrict our analysis to $\beta >0$.

Equations~(\ref{unified'}) and (\ref{madm}) corresponds to a {\it linear} form of the GUP.  However,  the conditions $r_{\rm CS}\rightarrow \RC$ for $M \ll \Mpl$ and $r_{\rm CS}\rightarrow\RS$ for $M \gg \Mpl$ could also hold for more complicated versions of the GUP.  For example, a {\it quadratic} form  that meets the asymptotic criteria corresponds to
\beq
M_{\rm ADM} \approx \sqrt{M^2 +  \beta^2 \Mpl^4/(4M^2)} \quad \Rightarrow \quad  r_{\rm CS} =  \sqrt{4M^2/\Mpl^4 +\beta^2 /M^2} \, .
\label{quadm}
\eeq
Such a relation has been shown to arise in Loop Quantum Gravity \cite{CMP11}.  In this case,  $\beta$ is the minimum area in units of $\ellp^2$; this is $4 \pi \sqrt{3} \gamma$ where $\gamma$ is the Immirzi parameter.  
The fact that the BHUP correspondence (whatever its form) allows black holes below the Planck mass  with a radius of order  $\RC$ rather than $\RS$ suggests a fundamental connection between elementary particles and black holes~\cite{CMP11,Carr16,CMN15}.  In contrast to the self-completeness picture, there is no longer a clear distinction between the two.  

Of course,  particles have charge and spin, so one has to go beyond the Schwarzschild case to  explore this possibility. Indeed, the extension of the $`M+1/M'$ proposal to the  Reissner-Nordström and Kerr cases has been studied in Ref.~\cite{CMMN20}.
In this case, there is an inner horizon and it is interesting that its radius for  Reissner-Nordström scales as $M^{-1}$ even in the standard classical solution.  However,  this only applies above the Planck mass and below the Planck length~\cite{CMMN20}, so one still needs to invoke the $`M+1/M'$ proposal to go below the Planck mass.
The issue of the BHUP correspondence and particle-like descriptions of microscopic black holes has been explored further in various contexts~\cite{FKN16,SpS21,Carr:2022ndy}.  
We also note that the similarity of the Compton wavelength and the gravitational scale 
suggests an alternative quantum setting, in which 
the BHUP correspondence implies a sub-Planckian dimensional reduction \cite{MuN13}. However, we do not explore this suggestion further here.

Since $ r_{\rm CS}$ has a minimum value of
$2 \sqrt{2\beta} \, \ellp$ and $\ellp$ 
 is the minimal possible length, we require $\beta >1/8$.  It makes no sense for $\beta$ to be less than this, since sub-Planckian smearing would then negate the effect of the GUP.  Since Eq.~\eqref{alphabeta} implies $\alpha \beta = 1/\epsilon^2$, this lower limit on $\beta$ implies an upper limit $\alpha < 8/ \epsilon^2$. 
On the other hand,  Eq.~\eqref{eq:gup} implies that $\Delta x$ has a minimum at $\Delta p = 1/\sqrt{2 \alpha} \, \Mpl$ and $\Delta x = \sqrt{2\alpha} \, \ellp$, so
a similar argument requires $\alpha >1/2$,  which implies $\beta < 2/\epsilon^2$ .  
One therefore requires $\alpha$ and $\beta$ to be in the range
\be
1/2 < \alpha < 8/\epsilon^2,  \quad 1/8 < \beta < 2 /\epsilon^2\, .
\label{range}
\ee
where we have argued that $\epsilon$ may be $1$.  In any case,  the limits in Eq.~\eqref{range} are only consistent for $\epsilon < 4$, which implies $\alpha \beta > 1/16$ and means that
the theoretically allowed range is much narrower than the experimental limits 
discussed below.
Note that the work of Buoninfante et al. ~\cite{Buoninfante_2019} on GUP stars assumes $\beta$ is close to one and points out that LISA could place strong limits on $\beta -1$.
Work by Bishop {\it et al. } \cite{Bishop:2023hvz} on the GUP in three dimension suggests $M+1/M$ may only be a leading term but implies $\alpha = \pi^2/8 \approx 1$.

\section{
Experimental Constraints on
$\alpha$ and $\beta$}
\label{sec:wfl}

Low-energy GUP
experiments include studies of atomic \cite{Das:2008kaa} and quantum optical systems \cite{Pikovski:2011zk} and composite particles \cite{Kumar:2019bnd}, as well as condensed matter systems \cite{Iorio:2017vtw} and macroscopic harmonic oscillator experiments \cite{bushev19,Bawaj:2014cda}. These provide an upper limit on $\alpha$ which is much larger than $1$. 
GUP can also be tested at high energies  ($M > \Mpl$) with gravitational bar detectors \cite{Marin:2013pga,Marin_2014} but this gives a very weak constraint. 
The GEH has been probed at high energies through direct and indirect probes of gravity in ``table top'' experiments \cite{WHP21,Bekenstein:2012yy,Bekenstein:2013ih},, 
with more theoretical effects being proposed in the context of gravitational decoherence \cite{Petruzziello:2020wkd}, astrophysical systems \cite{Moradpour:2019wpj}, and inflationary models
 \cite{Easther:2001fi}.   These provide an upper limit on $\beta$ which is much larger than $1$.  
GEH can also be tested at low energies ($M < \Mpl$) but without the modifications discussed in Sec.~4 this gives absurdly strong constraints, essentially requiring $\beta =0$.
In our `$M+1/M$' model,  one expects  $\beta \sim \alpha^{-1}$,  so
the GUP upper limit $\alpha _{\rm max} \gg 1$ becomes a lower limit $\beta _{\rm min} \ll 1$, which complements the GEH upper limit $\beta _{\rm max} \gg 1$. 
The permitted range for $\beta$ is compatible with the expected range around $1$ but much more extended.  One might also constrain $\beta$ by observations on astrophysical scales but in this domain the effects of the Extended Uncertainty Principle, in which $ \Delta x \Delta p \sim 1 + (\Delta x)^2$ rather than $1 + (\Delta p^2)$, becomes more relevant \cite{Mureika_2019}. 

\subsection{GUP tests for $M < \Mpl$}

Constraints on the GUP parameter $\alpha$
arise from a variety of mechanical oscillator experiments 
because corrections in the $[x,p]$ commutation relation result in a dependence of the resonant frequency of the oscillator on its amplitude.  
In particular,
Bushev {\it et al. }~\cite{bushev19} have measured this effect for a 0.3 kg ultra-high-Q sapphire split-bar mechanical resonator, using a $10^{-5}$ kg quartz bulk acoustic wave resonator, and find $\alpha< 5.2 \times 10^6$.  Experiments by Pikovski {\it et al. } \cite{Pikovski:2011zk} with quartz resonators in the mass range $10^{-11}-10^{-7}~$kg yield a stronger limit $\alpha< 4 \times 10^4$.  
Bawaj {\it et al.} \cite{Bawaj:2014cda} report oscillator-type experiments in the mass range $10^{-11}-10^{-4}~$kg, with corresponding limits on $\alpha$ in the large range $10^{26} -10^7$. If we assume $\beta = \alpha^{-1}$ from Eq.~\eqref{alphabeta} with $\epsilon = 1$, 
 these  upper bounds on $\alpha$ imply lower bounds 
$\beta > 2 \times 10^{-7}$ and $\beta > 3 \times 10^{-5}$, respectively\footnote{Bushev {\it et al.} and Bawaj {\it et al.} call the GUP parameter $\beta_0$; this is different from our GEH parameter $\beta$.}.

Data from the 1936 measurement of a pendulum period by Atkinson \cite{atkinson1936relation} could potentially lead to a much stronger limit.
However, due to the lack of a proper frequency stability measurement in these experiments, the exact upper bound cannot be reliably established.  Moreover,  pendulum-based systems only test a specific form of the modified commutator that depends on the mean value of the momentum.  Electro-mechanical oscillators, by contrast,  test any form of the GUP due to their greater stability and higher degree of control.

One can extract extremely weak upper limits to possible quantum gravity effects from spectroscopic
measurements in the hydrogen atom, whose energy levels are predicted very precisely. The
deduced upper limits are $\alpha < 10^{36}$ using the Lamb shift \cite{Das:2008kaa} and $\alpha < 4 \times 10^{34}$ using the 1s-2s level energy difference \cite{Quesne_2010}.  
There are also weak limits from  particle physics experiments at CERN because 
$\alpha$ is
the square of the number of Planck lengths
over which the position
 has been ``smeared''due to gravitational effects. 
Currently the LHC has probed length scales of order $10^{-17}~$cm,
which corresponds to 
$\sim 10^{16}\ell_{\rm P}$, so $\alpha$ 
can be at most $10^{32}$.  But both these limits are far weaker than the other constraints discussed below.

\subsection{GUP tests for $M > \Mpl$}

The GUP is usually associated with sub-Planckian energies  (i.e.  with the behavior of particles).  Indeed, it is the extrapolation to super-Planckian energies which brings in the link with black holes and motivates the BHUP correspondence.  This raises the issue of whether there can be any GUP tests in the $M > \Mpl$ regime.  However,  there is a semantic issue here since super-Planckian {\it masses} may be involved in an experiment even though the {\it energy} involved is sub-Planckian.  For example,  many authors have considered GUP modifications to Hawking radiation  even though these predictions have never been experimentally verified.  In this situation,  the black hole has $M > \Mpl$ but the emitted particles have sub-Planckian energy since the black hole is cooler than the Planck temperature ($T_{\rm P} \sim 10^{32}$K).  This is the sense in which the tests described in this subsection might be described as sub-Planckian. The tests in the previous subsection usually involve sub-Planckian masses, although the experiments of Bawaj {\it et al.} \cite{Bawaj:2014cda} span both $M < \Mpl$ and $M > \Mpl$.

Marin {\it et al.} \cite{Marin_2014} have shown that the very low mechanical energy measured in the main vibrational mode of gravitational-wave bar detectors can set an upper limit to $\alpha$. In particular, a bound is derived by exploiting the sub-milli-Kelvin cooling of the normal modes of the ton-scale gravitational-wave detector AURIGA to place an upper limit for possible Planck-scale modifications on the ground-state energy of an oscillator \cite{Marin:2013pga}.  A crucial issue is the appropriate value of the mass. The observed motion can be described by the equation of a damped harmonic oscillator with effective mass $M_{\rm eff}$.  But the deviations from standard quantum mechanics also depend on the probe mass. They argue that this implies a reduced mass which is half the mass of the bar.  The AURIGA experiment then leads to a limit $\alpha < 3 \times 10^{33}$, which is very weak but somewhat stronger than the Lamb shift limit mentioned above\footnote{Marin {\it et al.} call the GUP parameter $\beta_0$; 
they also use a
parameter $\beta \equiv [\alpha (\hbar m \omega_o/(\mpl^2c^2)]$, where 
$\omega_o$ is the oscillator frequency,  but this is different from our GEH parameter $\beta$.}.

\subsection{GEH tests for $M > \Mpl$}

In the 
GEH expression, $\beta$ represents the square of the number of Planck lengths
over which the position
 has been ``smeared', so in principle there could be quantum corrections to the event horizon radius in the super-Planckian regime.  These are currently unmeasurable but
constraints on the GEH parameter $\beta$ may be obtained from a variety of table-top experiments that probe 
Newtonian gravity at small length and mass scales. 
These generally involve a weak gravitational field and 
precision force measurements.  
From the perspective of General Relativity, the  
ADM mass in Eq.~\eqref{madm} will alter the spacetime structure according to Eq.~\eqref{newmetric}.
Since this can be written as
\beq
g_{00} = 1+2\Phi_{\rm N}(r)~,
\label{weak1}
\eeq
where $\Phi_{\rm N}(r) = -GM_{\rm ADM}/r$ is the Newtonian potential, . 
these experiments are sensitive to 
$\beta$ in both the super-Planckian and sub-Planckian regimes. 
Equation~\eqref{newmetric} implies a potential 
\beq
\Phi_{\rm GEH}(r) = -\frac{GM}{r} \left[ 1  + \frac{\beta}{2} \left( \frac {\Mpl}{M}\right) ^2 \right]  \, .
\label{gupnewt1}
\eeq
The second term in Eq.~\eqref{gupnewt1} becomes important for masses around
$\Mpl \sim 10^{-5}~$g, which is only four orders of magnitude smaller than those 
probed in the microgravity experiments discussed by Westphal {\it et al.}~\cite{WHP21}. We might therefore expect observable deviations from the Newtonian prediction even in the super-Planckian domain..
The fractional difference between the Newtonian and GUP potentials is
\beq
\frac{\Delta \Phi}{\Phi_{\rm N}} = \frac{|\Phi_{\rm N}(r)-\Phi_{\rm GEH}(r)|}{\Phi_{\rm N}(r)}= \frac{\beta}{2} \left( \frac {\Mpl}{M}\right) ^2 \, .
\label{sg}
\eeq
Setting $M=0.1~$g, the mass involved in the experiments described in Ref.~\cite{WHP21},
and requiring the fraction \eqref{sg} to be small then implies
\beq
\beta \ll \left(\frac{M}{\Mpl}\right)^2  \sim 10^8 \, .
\label{mc}
\eeq

The GEH parameter $\beta$ can be further constrained by 
considering the gravitational {\it force} between two masses.
Westphal {\it et al.}~\cite{WHP21} measured the gravitational force between two 90~mg masses with a centre-of-mass separation in the range 2.5 - 5.8~mm. 
The standard Newtonian gravitational potential generated by each mass is then 
$\Phi_{\rm N} \sim 2.4\times 10^{-12}~$J/kg and the force is $F \sim 10^{-13}~$N.  
For equal masses 
\if
$M_1 = M_2$
with separation $r_{12}$,  the modified force can be expressed as
\beq
F_{\rm GEH} = \frac{M M_{\rm ADM}}{\Mpl^2 r_{12}^2} \quad {\rm or} \quad \frac{M_{\rm ADM}^2}{\Mpl^2 r_{12}^2} \, ,
\eeq
depending on whether only the active gravitational mass or both the passive and active masses are modified.
\fi
the fractional change in the force is
\beq
\frac{\Delta F}{F} \sim \beta\left(\frac{\Mpl}{M}\right)^2 \quad {\rm or} \quad \beta^2 \left(\frac{\Mpl}{M}\right)^4 \, ,
\label{Fchange}
\eeq
depending on whether only the active gravitational mass or both the passive and active masses are modified. Given the 
experimental value of $G$, which is
$(6.04\pm0.06)\times 10^{-11}\rm N \,
m/kg^2$, we can express the relative force uncertainty as 
\beq 
\frac{\Delta F}{F} < \frac{\Delta G}{G} \sim 10^{-2} \, .
\eeq
Equation~\eqref{Fchange} then implies $\beta < 10^6$ or $10^7$, somewhat stronger than limit \eqref{mc}. 
The error is dominated by the distance uncertainty, which varies from 7\% to 16\%.  Since the distance enters the force quadratically,  this implies an error on $F$ of 14 - 32\%, so the limit on beta is of order $10^7$ and $3 \times 10^7$ in the linear and quadratic cases, respectively.

\subsection{GEH constraints for $M < \Mpl$}

In this subsection, we will discuss theoretical constraints, required to avoid disrupting standard physics in the $M < \Mpl$ regime,  rather than experimental tests.  However, these constraints are crucial since they raise a fundamental problem with the $M+1/M$ model whose possible resolutions are addressed in Section 4.  For
$M < \sqrt{\beta} \Mpl$, the second term in the potential (\ref{gupnewt1}) dominates and gives
\beq
\Phi (r) \approx - \frac{\beta}{2Mr}  \approx - \frac{\beta \RC}{2r}~,
\label{phinew}
\eeq 
where $\RC$ is the standard Compton wavelength for the source mass $M$.
The potential is small (i.e.  much less than $1$) providing $r \gg \beta \RC$,  which is the usual Compton condition apart from the factor $\beta$. 
For example,  the potential due to the 
proton in an atom
at the Bohr radius, $r_B \sim \alpha_e^{-1} m_e^{-1}$,  is small providing
\beq
\beta \ll \frac{r_{\rm B}}{\RC} \sim \alpha_e^{-1}\left(\frac{m_{\rm p}}{m_e}\right) \sim 10^{5}  \, ,
\eeq
where $\alpha_e = e^2/\epsilon_o \hbar c \sim 10^{-2}$ is the electric fine structure constant.  
The Compton wavelength for a proton is $\RC \sim 10^{-15}$m and the Bohr radius $r_B \sim 10^{-10}$m.

Determining the {\it force} between 
the electron and proton in the Bohr atom is less straightforward.  Assuming both the active and active gravitational masses are affected,  the force is 
\beq
F_{\rm GEH} =
\frac{m_{\rm p} m_e}{\Mpl^2 r_B^2}\left(1+\frac{\beta}{2}\frac{\Mpl^2}{m_{\rm p}^2}\right)\left(1+\frac{\beta}{2}\frac{\Mpl^2}{m_e^2}\right) \, .
\label{FGEH}
\eeq
Providing 
\beq
 \beta > \left( \frac{m_{\rm p}}{\Mpl}\right)^2 \sim 10^{-38},
\label{betamin}
 \eeq
both factors of $1$ in Eq.~\ref{FGEH} can be beglected,  so it gives
\beq
F_{\rm GEH} \approx  \frac{\beta^2\Mpl^2}{4m_{\rm p} m_e r_B^2} \, .
\eeq
This is less than the electric force between the electron and proton, ensuring the stability of the Bohr atom, only if
\beq
\beta < e \, (m_{\rm p} m_e)^{1/2} /\Mpl \sim 10^{-21} \, .
\eeq
This 
 is certainly inconsistent with the limit $\beta > 1/8$ from Eq.~\eqref{range}.  However, if only the active gravitational mass is affected, the limit becomes 
\beq
\beta < e^2 (m_{\rm p}/m_e) \sim 10 \quad {\rm or} \quad \beta < e^2 (m_e/m_{\rm p}) \sim 10^{-5}  \,,
\eeq
depending on whether the proton or electron is regarded as active.  Only the first condition is consistent with Eq.~\eqref{range}.

For two protons at distance $r$, we can regard the extra force for $M < \Mpl$ as corresponding to a `strong gravity' effect,
\beq
F_{\rm GEH}= \frac{G_{\rm eff} m_{\rm p}^2}{r^2} \,,
\eeq
where the effective gravitational constant is given by 
\bea
\frac{G_{\rm eff}}{G} =
\begin{cases}
 \beta^2 (\Mpl/m_{\rm p})^4  \sim 10^{76} \beta^2  & ({\rm passive \; and \; active \; mass \; affected})\\
 \beta (\Mpl/m_{\rm p})^2  \sim 10^{38} \beta   & ({\rm just \; active \; mass \; affected}) \, .
\end{cases}
\eea 
(Unlike the usual strong gravity force, this not a short-distance effect.) If we require the gravitational force to be less than the electric force between the two protons,  which is $10^{38}$ times the Newtonian force,  this requires 
\beq
\beta < e \, (m_{\rm p}/\Mpl) \sim 10^{-20} \quad {\rm or} \quad \beta < e^2  \sim 10^{-2}  \,.
\eeq
However, both these requirements are inconsistent with 
Eq.~\eqref{range}.  Of course,  the constraint from the force between two electrons would be even stronger. 

\section{Possible modification of M+1/M model} 
\label{sec5}

In this section, we address the question of whether the large extra force predicted by the `$M+1/M$' model at small mass scales can be obviated or reinterpreted in some way.  This is not only motivated by a desire to avoid the constraints on $\beta$ discussed in Section 3.4.  It is also prompted by various conceptual problems we have encountered.  For example,  we have seen that there is an ambiguity in whether the active or passive gravitational  mass -- or both -- are affected by the `$M+1/M$' term. This makes a huge difference to the constraints discussed in Section 3.4 and it also affects the status of Newton's 3rd law.  
Another problem concerns the gravitational field of a composite object.  If the active gravitational mass is $M$,  as in the standard case,  this is just the sum of the fields of the components. But the addition of the $1/M$ term violates this condition since $\Sigma (1/M_i) \neq 1/ (\Sigma M_i)$.  

One way around these problems is to assume that the gravitating mass is just $M$ but in this case why does the gravitational radius scale as $M+1/M$? 
A more radical possibility is
that Newtonian gravity does not operate at all at sub-Planckian mass scales. Indeed, it is tempting to appeal to  the historical tension between classical and quantum predictions in this context.  For example,  classical mechanics predicts the rapid collapse of the atom and only the introduction of quantum mechanics ({\it i.e.} the semi-classical Bohr model and ultimately the Schr\"odinger equation) can avoid this.  So if one has to choose between Newtonian theory and the GUP model – which is itself a prediction of some quantum gravity models – it might seem natural to assume that the latter is a more fundamental tenet of physics.   

In the first subsection below,  we discuss the general issue of the gravitational force between elementary particles, raising the possibility that the $1/M$ force -- and perhaps even the $M$ force -- may not operate at all (cf.  the asymptotic freedom between quarks under the strong force).   But a less radical approach is to accept the existence and naive interpretation of the $M + 1/M$ term and then try to restore consistency in some way.  This motivates the various avenues of resolution discussed in the following subsections.

\subsection{Gravitational force between elementary particles}

The constraints in the previous section on
$\beta$ on microscopic scales 
assume that there is a standard Newtonian gravitational force between particles, albeit with a modifed expression for $M$.  However, although we are worried about the effect of the $1/M$ term,
we cannot even be sure of the $M$ term in this context.  Since Newton's law has only been confirmed for masses down to  $0.1$~g, there is no direct evidence that it applies for elementary particles at all.  Either the inverse-square law or mass-product rule may fail and, in principle, 
experiments could have detected both of these effects.
However, it is important to note that recent experiments have demonstrated that antimatter falls under gravity in the same way as matter~\cite{anderson}.  The possibility that there is no gravitational force on sufficiently small mass and/or length scales is reminiscent of  asymptotic freedom for quarks \cite{Gross} and the possibility that gravity is replaced and/or enhanced by some other force is reminiscent of  the strong gravity proposal \cite{Sivaram:1975dt}.  Whatever the case, the transition would most naturally occur at (or at least very near the Planck mass. 

There is also the issue of what is meant by $M$ for an elementary particle since one cannot define this in gravitational terms if Newton's law itself is suspect. In particular,  the assumption that the active and massive gravitational masses are the same -- as required by the equivalence principle on macroscopic scales -- becomes questionable  in this case.  One possibility would be to associate the mass with internal energy but this is also suspect. For example, the classical radius of the electron identifies its rest mass with its electrostatic energy ($r_{\rm e} \sim e^2/m_{\rm e} \sim 10^{-13}$cm) but this is smaller than its Compton wavelength by a factor of $\alpha_{\rm e}$,  so the electrostatic energy is smaller than the electron rest mass ($m_{\rm e} \sim 10^{-27}$g) by the same factor.  Coincidentally,  the classical size of the electron is roughly the Compton wavelength of the proton and the classical size of a proton ($r_{\rm e} \sim e^2/m_{\rm p}\sim 10^{-16}$cm) is roughly the Compton wavelength of a W-particle.  The rest mass of the proton  ($m_{\rm p} \sim 10^{-24}$g) is far more than the sum of the rest masses of its constitutent quarks because one must also allow for the gluon energy and the kinetic energy of the quarks,  both of which  are uncertain.

Since the essential issue is how to incorporate gravitational effects into quantum theory, we should also recall the role of the Schr{\"o}dinger-Newton (SN) equation. This is a modification to the Schr{\"o}dinger equation suggested by Di{\'o}si \cite{ Diosi:1984wuz} and Penrose \cite{ Penrose2014-PENOTG-2} to take into account the gravitational potential generated by the distribution of mass in a system.  In particular, Jusufi and Ali \cite{Jusufi:2023tdn} apply this to calculate the regularized self-energy of a particle, using the form of the gravitational and electrostatic potentials derived from string T-duality.  
This results in a correction to the energy of the particle.
They then extend their analysis to relativistic particles using the Klein-Gordon, Proca and Dirac equations and find that the corrected energy leads to a form of the GUP which depends on the spin of particle.  Although their results differ from ours in some respects, it is interesting that they obtain different  forms of the GUP for bosons and fermions.

\subsection{Association of extra force with electric interactions}

If the $1/M$ term only affects the {\it active} gravitational mass, the `extra' gravitational force between two objects of mass $M$ associated with the $1/M$ term is
\bea
\Delta F_{\rm G} = 
\frac {G \beta \mpl^2}{2r^2} \, ,
\eea
with no dependence on $M$.  [If the $1/M$ term also affects the {\it passive} gravitational mass,  $\Delta F_{\rm G}$ would be increased by another factor of $(\Mpl/M)^{2}$, so the extra force is much larger and also depends upon $M$, but we assume this is not the case in this subsection.]  If the objects have charge $ne$, the ratio of $\Delta F_{\rm G}$ to the electrostatic force between them is
\bea
\frac{\Delta F_{\bf G}}{F_{\bf E}} = \frac{G \epsilon_o \beta \mpl^2}{2 n^2 e^2} = \frac{\beta}{2 \alpha_{\rm e} n^2}
\eea
For two protons or two electrons, $n=1$ and so one requires $\beta \ll  \alpha_e$ if one is not to disturb the measured electrostatic force between these particles.  

A more radical possibility is that the extra gravitational force {\it is} the electric force, since this avoids the possibility that the former dominates the latter.  For protons or electrons, or indeed any particle with $n=1$, this requires
\bea
\beta = 2 \alpha_{\rm e} \, ,
\label{alphabetalink}
\eea
although this goes against the constraint $\beta > 1/8$ of Eq.~\eqref{range}.  In this case, the electrostatic interaction in some sense {\it generates} the GUP effect.  However,  there are several problems with this suggestion: (i) What is regarded as the active gravitational mass in evaluating the electrical force between an electron and proton (eg.  in determining the Bohr radius), this giving an ambiguity of $(m_{\rm p}/m_{\rm e})^2 \sim 10^6$?
(ii) For positive $\beta$,  the extra gravitational forces should be attractive,  so why do similar charges (pp and ee) repel and opposite charges (ep) attract? (iii) Why do neutral particles have no extra gravitational force?
The resolution of (ii) and (iii) may be that the GUP parameter depends on the particle charge.
Thus $\alpha = 0$ for neutral particles and is negative for oppositely charged particles. This is not the usual GUP scenario but there is nothing to preclude it. Also we note that the black hole radius depends on charge as well as mass, so such a dependence may be in the spirit of the BHUP correspondence.  The resolution of (i) remains unclear.

The above discussion suggests that no
single values of $\alpha$ or $\beta$ may suffice
to explain all possible particle couplings via gravity. We posit instead that each particle species has
its own unique values for these parameters.
For example,  if only the active gravitational mass is modified, the force between an electron and proton can be the same (as required by Newton's 3rd law) if
\bea
\beta_{\rm e}/\beta_{\rm p}  = (m_{\rm e}/m_{\rm p})^2 \, .
\eea
More generally,  one might require $\beta \propto m^2$.  However, in this case,  Eq.~\eqref{alphabetalink} no longer applies.

Such a proposal is reminiscent of the equivalence-violation scenarios 
considered for neutrinos in the 1990s \cite{Mureika:1995ap,Mureika:1996de,Mureika:1996ud}. 
Along these lines, a recent paper \cite{Ali:2022ckm} suggests 
that the GUP provides 
an effective running of Planck's constant:
\beq
\Delta x \Delta p \geq \frac{\hbar^\prime}{2} \quad {\rm with} \quad \hbar^\prime = \hbar(1+2 \alpha\Delta p^2) 
= \hbar(1+2 \beta^{-1} \epsilon^{-2}\Delta p^2) \, ,
\label{varhbar}
\eeq
where we have used Eq.~\eqref{range} in the last step.  This argument also appears in Ref.~\cite{Marin_2014}. This was motivated by noting that the charge radius $r_*$ for a variety of particles hints at  
different values of Planck's constant, $\hbar^{\prime} = r_*mc$. 
Such an expression 
resembles the usual
 relationship between the Planck scales, $\hbar = \ellp \mpl c$ with the identifications $\ellp \rightarrow r_*,  \mpl \rightarrow m, \hbar \rightarrow \hbar^\prime$. 
Regarding the variable $\hbar^\prime$ as a running 
constant may also provide a solution to the cosmological constant problem~\cite{elmashad}.
However, this is only possible  if 
the value of $\beta$ 
is different for each particle.

\subsection{Association of extra force with strong interactions}

Since the strong coupling constant is $\alpha_S \sim 10$, one can associate the extra gravitational force with nuclear interactions if Eq.~(\ref{alphabetalink}) is replaced by 
\bea
\beta = 2 \alpha_{\rm S} \, ,
\label{alphaSbeta}
\eea
which is marginally consistent with the constraint $\beta < 2/\epsilon^2$ from Eq.~\eqref{range}.  This suggests that nuclear particles may also be associated with a GUP parameter (even if they have no electrical charge).  Indeed,  there might be a value of $\alpha$ for any conserved quantity, with the total value of $\alpha$ being additive.  One might also include angular momentum here,  in which case there would also be a contribution to $\alpha$ associated with spin. Again, this would be in the spirit of the BHUP correspondence since the black hole radius depends on angular momentum \cite{Ali:2022ckm}. 
If the GUP parameter $\beta$ is dependent on spin and therefore different for fermions and bosons, this would require spin to be a gravitationally sensitive characteristic of the particles. This could indeed be the case.  Although the traditional view of spin is that it is a purely quantum feature with no classical analog, Ohanian \cite{1986AmJPh..54..500O} published a review in 1985 of a derivation by Belifante \cite{1939Phy.....6..887B} that directly links spin to the rotational energy flow in the associated quantum field. In this case, spin is a direct product of energy and should gravitate, so particles of different spin should be affected differently by gravity.

This proposal is reminiscent of the strong gravity model of the 1970s except that this was assumed to be associated with a short-range force,  decreasing with distance as  $e^{-m_{\pi}r}$ where $m_{\pi}$ is the pion mass~\cite{Sivaram:1975dt}.  Such a term  would  not eliminate the extra force between particles at small distances but one way of doing so would be to add an $e^{-1/(m_{\bf c} r)}$ term for some mass $m_{\bf c}$.  Another approach would be to add an $e^{-m_{\bf c}/M}$ factor to the $1/M$ term, which cuts it off  below some mass $m_{\bf c}$ for all $r$ but leaves the $M$ term intact.  These possibilities correspond to a force of the form
\be
F(M,r) \sim \quad \frac{GM}{r} \left[1 + e^{-1/(m_{\bf c}r)} \frac{\beta}{GM^2}\right]  \quad {\rm or} \quad  \frac{GM}{r} \left[1 + e^{-m_{\bf c}/M} \frac{\beta}{GM^2}\right] \, .
\label{strong}
\ee
Both alternatives might seem somewhat arbitrary in the current context but the first one may be justified if we combine this approach with the high-dimensional model, so we return to this issue in Section 4.6.

\subsection{Association of extra force with irreducible mass}

The possible dependence of $\alpha$ on charge and spin prompts a reconsideration of what is meant by mass.
The irreducible mass $M_{\rm irr}$ for Kerr-Newman is defined as~\cite{Ruffini}
\be
 M_{\rm irr} = \frac{1}{2} (2M^2 - Q^2/G + 2M \sqrt{M^2 - Q^2/G - J^2/G^2 M^2} \,)^{1/2} \, . 
\ee
(where $c = 4 \pi \epsilon_0 = 1$) or equivalently
\be
 M_{\rm irr} = \frac{1}{2G}[ (GM  + \sqrt{G^2M^2 - GQ^2 - a^2} \,)^2 +  a^2 ]^{1/2} \, 
\ee
for a black hole with charge $Q$ in units of $\sqrt{\hbar \, c}$) and angular momentum $J = aM$ (in units of $\hbar$).  
This can be interpreted as the mass measured at future null infinity and it is also related to the black hole area
\be
A = 16 \pi G^2 M_{\rm irr}^2 = 8 \pi GM [GM - Q^2/(2M) + \sqrt{G^2M^2 - a^2 - GQ^2} \, ] \, .
\ee
In the $J=0$ case,  the irreducible mass becomes
\be
M_{\rm irr} = \frac{1}{2}(M + \sqrt{M^2 - Q^2/G} \, ) = \frac{1}{2G} r_+  \, ,
\label{irr}
\ee
so the horizon size is just the Schwarzschild radius associated with this mass.  In the $Q=0$ case,  the irreducible mass becomes
\be
M_{\rm irr} = \frac{M}{\sqrt{2}}[1 + \sqrt{1 - J^2/(G^2 M^4)}]^{1/2} = \frac{1}{2 G}( r_+^2 + a^2)^{1/2} \, .
\ee
As expected, both expressions reduce to $M$ in the Schwarzschild case. 

These expressions can be inverted to give $M$ in terms of $M_{\rm irr}$:
\be
M =M_{\rm irr} + \frac{Q^2}{4GM_{\rm irr}}
\label{RN'}
\ee
for Reissner-Nordstrom (RN) and
\be
M =\left( M_{\rm irr}^2 + \frac{J^2}{4 G^2 M_{\rm irr}^2} \right)^{1/2}
\label{K'}
\ee
for Kerr.  
Since $M$ is equivalent to the ADM mass (i.e.  the mass measured at spacelike infinity) in both the RN and Kerr solutions, it is interesting that Eqs.~\eqref{RN'} and \eqref{K'} resemble the expressions for the linear and quadratic versions of the GUP providing $M$ in Eqs.~\eqref{madm} and \eqref{quadm} is interpreted as $M_{\rm irr}$.  This suggests that there may be some connection beween the GUP and the notion of irreducible mass. 

On the other hand, the expression for the irreducible mass is entirely classical, so neither $M$ nor $M_{\rm irr}$ can be appreciably less than $\mpl$ and the size of the event horizon retains the standard form.
To allow the concept of a sub-Plankian black hole,  one needs to introduce the $M+1/M$ procedure itself.  
In the RN case, Eq.~\eqref{irr} is replaced by
\be
M_{\rm irr} = \frac{1}{2} \left( M + \frac{\beta}{2GM} + \sqrt{M^2 + \frac{\beta^2}{4 G^2M^2} + \frac{\beta}{G}  - \frac{Q^2}{G}} \, \right) 
= \frac{1}{2G} r_+  \, .
\ee
In the super and sub-Planckian limits this gives
\be
 r_+ \approx
\begin{cases}
G(M + \sqrt{M^2 - Q^2/G} \, )  & (M \gg  \sqrt{\beta} \Mpl) \\
\frac{\beta}{M} \left[ 1- \left( \frac{QM}{\beta \mpl} \right)^2 \right] & (M \ll \sqrt{\beta}  \Mpl ) \,  .  
\end{cases}
\ee
This is the standard RN expression in the super-Planckian regime and close to the Compton expression in the sub-Planckian regime.  The last term in the latter context can be neglected since $Q < M/\mpl$ and $M^2 \ll \beta \mpl^2$. A similar analysis and conclusion applies in the Kerr case.

Since both $Q$ and $J$ are zero in the context of our original $M+1/M$ proposal,  the significance of 
this analysis for that proposal is unclear.   However, it is interesting that a similar feature arises if one calculates the irreducible mass for the holographic metric, a quantum gravity improved Schwarzschild metric, that  has been proposed in the context of the gravity-self-completeness paradigm \cite{NiS14}. This has
\be
g_{00} = 1 - \frac{2 G  M_{\rm ADM}}{r} \left(\frac{ r^2   }{r^2 + \ellp^2}\right) \, ,
\label{holmet}
\ee
which implies
\be
M_{\rm irr} = \frac{1}{2G} r_+ = \frac{1}{2}(M_{\rm ADM} + \sqrt{M_{\rm ADM}^2 - \Mpl^2} \, ) \, .
\label{holM}
\ee
%where $M$  is again interpreted as the ADM mass. 
This can be inverted
to give
\be
M_{\rm ADM} =M_{\rm irr} + \frac{\Mpl^2}{4M_{\rm irr}} \, .
\label{invert}
\ee
The metric in Eq.~\eqref{holmet} results from a non-vanishing mass-energy distribution $\rho(r)$ centered around the origin. The cumulative mass distribution 
$m(r)$, which appears in $g_{00} =1-2G m(r)/r$, is given by
\be
m(r) = 4 \pi \int_0^r R^2 \rho(R) = M_{\rm ADM} r^2/(r^2 + \ellp^2) \, .
\ee 
The condition $g_{00}=0$ determines the event horizon radius $r_+$ as a function of $M_\mathrm{ADM}$. 
By inverting this relation, the ADM mass can be expressed in terms of the event horizon radius. For the Schwarzschild case,  one has $M_{\rm irr}=m(r_+)=M_\mathrm{ADM}$.
For the holographic case,  
$M_{\rm irr}$ and $m(r_+)$ still coincide from Eq. \eqref{holmet},  so 
the irreducible mass is still the
mass inside the event horizon, but they both
differ from $M_{\rm ADM}$:
\be
M_{\rm ADM}=\frac{r_+}{2G}\left(1+\frac{\ellp^2}{r_+^2}\right)> M_{\rm irr}=m(r_+).
\ee 
The amount of mass leaking outside the horizon corresponds to the aforementioned quantum atmosphere \cite{DvG12, DvG13,DvG13+,tGR+18,Gid17}.

Comparison of Eqs.~\eqref{invert} and \eqref{RN'} shows that the holographic metric has the horizon structure of RN with $Q=1$, so it is the quantum gravity analogue.   
Since $M_{\rm irr}$ for RN represents the bare mass, i.e.  what is left of the initial $M_{\rm ADM}$ after extracting the electromagnetic term with a Penrose process, for the holographic metric it represents the classical mass, without any contribution from quantum gravity self-interaction.  One can say that the total mass at infinity  $M_{\rm ADM}$ has two contributions: the bare term $M_{\rm irr}$ (dominant for super-Planckian black holes) and the quantum gravity self-interaction term $\mpl^2/2M_{\rm irr}$  (dominant for sub-Planckian black holes). Without further modifications, 
both $M_\mathrm{irr}$ and $M_{\rm ADM}$
for the holographic metric are bounded from below: $M_{\rm irr}\geq M_\mathrm{P}/2$ and $M_{\rm ADM}\geq M_\mathrm{P}$.  However, as in the RN case, we must adopt the $M+1/M$ procedure to satisfy the BHUP correspondence. Therefore Eq.~\eqref{holmet} cannot be the final quantum-gravity solution and one must 
extend the validity of Eq. ~\eqref{invert} to the full sub-Planckian regime,
so that it describes particles-black-hole systems (aka quantum mechanical black holes) whose $M_{\rm ADM}$ is dominated by quantum effects and goes like $M_{\rm irr}^{-1}$.

The concept of quasi-local energy $E(r)$ may also be relevant here since this approaches the ADM mass $M$ in the limit $r \rightarrow \infty$.  In the limit of zero angular momentum it becomes \cite{Martinez:1994ja}  
\be
E(r) = r - r \sqrt{1 - \frac{2M}{r}} \, .
\ee
At the event horizon $r_+$ this gives
\be
E(r_+) \approx r_+ \left[ 1 + \frac{a^2}{2r_+^2}\right] \approx 2M \left[ 1 + \frac{a^2}{8G^2 M^2}\right] \approx 2 M_{\rm irr}
\ee
where the last expression applies in the slow-rotation regime.

In all these cases, the question arises of how one eliminates the large extra gravitational force between particles with sub-Planckian mass? One way would be to associate the gravitational force with the irreducible mass, since this would remove the extra term altogether.  Indeed,  one might try to justify this on the grounds that $M_{\rm irr}$ is the mass measured at future null infinity. 
However,  in this case the outer horizon radius is no longer the Compton scale in the sub-Planckian range and this  
goes against the spirit of the `$M+1/M$' model.  
In the next subsection, we consider a ``two-space" interpretation of the metric which may resolve this problem.

\subsection{Association of extra force with other spaces}

In the Loop Quantum Gravity (LQG) scenario,  a quadratic form for the GUP arises  because the circumferential radius $R$ in the modified Schwarzschild metric  is related to the radial coordinate $r$ by 
\be
R = (r^2 + a_0^2/r^2)^{1/2}
\ee
where $a_0$ is the minimum area in LQG. This replaces the usual central singularity at $r=0$ with a wormhole which connects to another asymptotically-flat space ($r \rightarrow 0 \Rightarrow  R \rightarrow \infty$).  The black hole has outer and inner horizons at
\be
r_+ = 2GM  , \quad r_- = 2GMP^2 \, ,
\ee
where $P \ll 1$ is the polymeric function.  The value of $R$ at the outer horizon is
\be
R_{\rm S} = 2G[M^2 + \beta^2 /(G^2M^2)]^{1/2} \equiv 2 G M_{\rm ADM} \, 
\label{LBH}
\ee
with $\beta = a_0/(4G)$ and the definition of $M_{\rm ADM}$ being different from Eq.~\eqref{quadm}.  So this  corresponds to the quadratic form of the BHUP correspondence.
The inner horizon is on the other side of the wormhole throat ($r = a_0^{1/2}$) and appears as an outer horizon for a black hole of mass $\Mpl^2/M$ in the other space.   The fact that a purely geometrical condition in LQG implies the GUP suggests some deep connection between general relativity and quantum theory. 

The important point in the present context is that the asymptotic (ADM) mass in our space is $M(1+P)^2 \approx M$, so there is no extra gravitational force associated with the $1/M$ term in Eq.~(\ref{LBH}).  In some sense, the extra force has been pushed into the other space,  where the black hole mass appears to be $\Mpl^2/M$. So perhaps there is an analogous solution (not necessarily associated with LQG) in which the circumferential radius in the modified Schwazschild metric is 
\be
R = r + a_0/r \, .
\label{linR}
\ee
This corresponds to our `$M+1/M$' model but without the problem of an extra gravitational force in our space.  
We note that the simplest Morris-Thorne wormhole \cite{1988PhRvL..61.1446M} has a circumferential radius
\be
R = \sqrt{r ^2 + b^2} \approx r \left( 1 + \frac{b^2}{2r^2} \right) = r + \frac{b^2}{2r} \quad {\rm for} \quad r \gg b \, ,
\ee
which resembles Eq.~(\ref{linR}). In any case, this raises the possibility that the $r_+$ may have the Schwarzschild form, even though the physical horizon radius is modified.

This suggests that it may be possible to combine the 2-space and irreducible mass approaches.  Both of these avoid an anomalous force between particles in {\it our} space but the latter does not allow the {\it outer} horizon to be associated with the Compton wavelength. However,  in the 2-space approach, the inner horizon of the black hole in our space becomes the outer horizon in the other space.

\subsection{Higher dimensional model}

\begin{figure}[h]
\includegraphics[scale=.33]{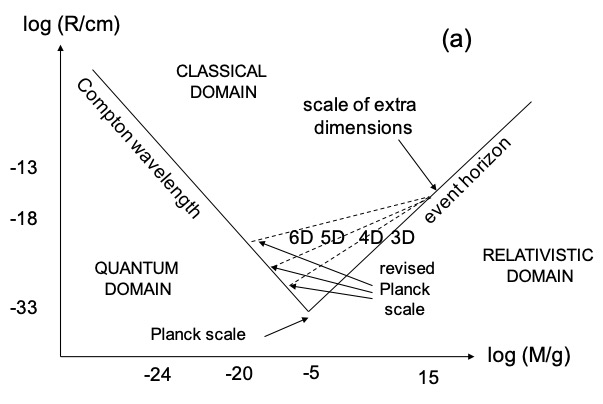}
\includegraphics[scale=.33]{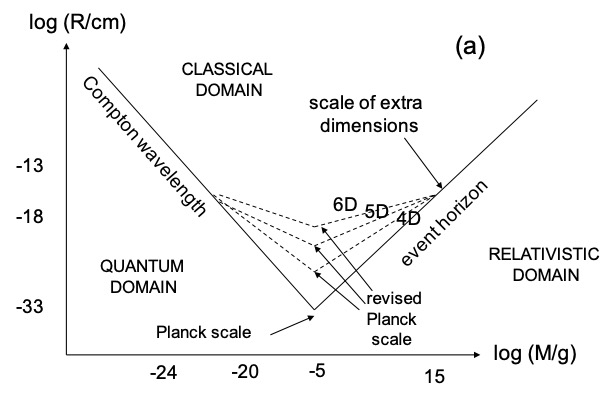}
\caption{Compton and Schwarzschild scales in the higher-dimensional regime, assuming the standard (a) and dual (b) expressions for the Compton wavelength.
}
 \label{highdim}       
\end{figure}

If there are $n$ extra spatial dimensions, all compactified on some scale $R_{\bf c}$, then the form of the unified expression $r_{\rm CS}(M)$ will be modified below $R_{\bf c}$  For black holes on the super-Planckian side,  the Schwarzschild radius $r_{\rm S}$ scales as $M^{1/(1+n)}$ for $\Mpl < M < M_{\bf c} \equiv R_{\bf c}/(2G)$.  For particles on the sub-Planckian side,  the Compton wavelength $r_{\rm C}$ scales as  $M^{-1/(1+n)}$ or $M^{-1}$ for  $\Mpl > M > \Mpl^2/M_{\bf c}$, depending on whether or not one preserves duality between the Schwarzschild and Compton expressions \cite{LaC18}. Both dependences on $M$  always increase the Planck length and they also decreases the Planck mass in the first case (allowing TeV quantum gravity and black hole production in accelerators).  
The form of $r_{\rm CS}(M)$ for the two cases and different values of $n$ is illustrated in Figure \ref{highdim}.  

It is not clear how the `$M+1/M$' model and even the concept of the ADM mass should be extrapolated to extra dimensions. The simplest model involves a potential of the form
\be
V(r) \propto  \frac{M_{\rm ADM} }{r^{1+n}}  \quad {\rm with} \quad M_{\rm ADM} \propto \left(M + \frac{1}{M} \right) \, .
\label{dual}
\ee
Assuming the black hole horizon corresponds to $V = 1$, this implies that the higher-dimensional Schwarzschild radius is
\be
r_{\rm S} \propto  M_{\rm ADM}^{1/(n+1)} \propto  \left(M + \frac{1}{M} \right)^{1/(1+n)} \, .
\label{HDS}
\ee
This reduces to the usual higher-dimensional form $r_{\rm S} \propto M^{1/(1+n)}$ for $M_c \gg M \gg \Mpl$ and it gives a Compton scale $r_{\rm C} \propto M^{-1/(1+n)}$ for $\Mpl^2/M_c \ll M \ll \Mpl$, which supports the duality proposal. 
However,  this is not the only possibility since the same asympototic forms would apply if
\be
r_{\rm S} \propto  M^{-1/(n+1)} + M^{1/(1+n)}  \, \Rightarrow  
\,  M_{\rm ADM} = M [1 + (M/\mpl)^{-2/(1+n)} ]^{1+n} \, .
\ee
On the other hand,  the standard (duality-violating) result, $r_{\rm C} \propto 1/M$, requires
\be
r_{\rm S} \propto  M^{1/(1+n)} + \frac{\mpl^{(2+n)/(1+n)}}{M} \, \Rightarrow \, M_{\rm ADM} = M [1 + (M/\mpl)^{-(2+n)/(1+n)} ]^{1+n} \, ,
\ee
which corresponds to a potential of the form
\be
V(r)  \propto \frac{M}{r^{n+1}} \left( 1 + (M/\mpl)^{-\frac{2+n}{1+n}} \right)^{n+1} \,.
\label{nondual}
\ee
Both cases give $M_{\rm ADM} \sim M+\mpl^2/M$ in the $n=0$ case. 
Perhaps Eq.~\eqref{dual} is more natural than Eq.~\eqref{nondual}
but it is clear that the higher-dimensional extension of GUP is not unique.  The forms proposed in Refs.~\cite{LaC18} and \cite{Cas13}  imply a softer degree 
of ultraviolet convergence of integrals in momentum space, making gravity weaker as the number of dimensions increases. This has been noted in Ref.~\cite{Knipfer:2019pgi} where a universal momentum space convergence for the GUP in any dimension has been proposed. 

Neither potential (\ref{dual}) or (\ref{nondual}) resolves the extra force problem.  However,  there is a natural link with models 4.2 and 4.3, since the main motivation for extra dimensions is to explain non-gravitational forces, so the combination of these approaches  may offer a solution.
In Kaluza-Klein theory, the electric interaction
is associated with a compactified 5th dimension,  so model 4.2 -- which links the electric force with the GUP parameter -- would suggest that the GUP is 
itself associated with an extra dimension.  
This raises the question of whether higher dimensions could not only {\it modify} the GUP effect but also {\it generate} it. The generalization of Kaluza-Klein theory to other forces suggests an extension of this approach to still higher dimensions.  This is especially interesting if there is a hierarchy of extra dimensions with increasing compactification scales \cite{2013MPLA...2840011C}.  As argued by Witten \cite{1981NuPhB.186..412W} in the context of SU(3)$\times$SU(2)$\times$U(1), the weak and strong interactions require two and four extra dimensions, respectively. This gives a total of 11 dimensions,  as predicted in M-theory \cite{1995NuPhB.443...85W} and discussed further in a footnote in Ref.~\cite{aurilia}{\footnote{Although we are here advocating a positive value of $n$,  Ref. ~\cite{CMN15} points out that $r_{\rm S}$ has the Compton form if the number of spatial dimensions is reduced to $1$, corresponding to $n=-2$. }.

So the brane and $M+1/M$ scenarios may be mutually beneficial: the extra dimensions may identify the $1/M$ term with known non-gravitational forces, while the compactification scale $R_{\rm c}$ may provide the exponential term in the first expression in Eq.~\eqref{strong}. We have not provided a specific model for such a cut-off but at least such a scale arises in principle.  It may also be relevant that the Reissner-Nordstrom metric contains $M/r$ and $Q/r^2$ factors, since the latter $r$-dependence corresponds to a 5-dimensional black hole. 
There is a large literature on black holes in Kaluza-Klein theory \cite{1985JMP....26.2308M,1986AnPhy.167..201G,
1995PhRvL..75.4165C} but  it is interesting that the charged 5D Kaluza-Klein black hole has \cite{Briet:2008mz}
\bea
g_{tt} = 1 - 2M/r + G Q^2/r^2 \, ,   \quad   g_{rr} = (1 - 2M/r + Q^2/2Mr)^{-1}
\eea
and the factor $M - Q^2/4GM$ (cf. the irreducible mass) appears in most equations.  
Apart from the sign, this is equivalent to Eq.~\eqref{madm} with $\beta = Q^2/2$. 

\section{Conclusions and relating these approaches}
\label{concl}

In this paper,  we have pointed out a serious problem with our `$M+1/M$' model: the anomalous gravitational interaction between sub-Planckian objects (i.e.  all known elementary particles). We have explored various ways of resolving this problem but without indicating a preferred option.  Therefore,  in concluding, we first summarize the advantages and disadvantages of each approach.\\

{\it Model 4.2}. Associating the extra  force with the electric interaction has the advantage that the force between identical particles is independent of mass and falls as $1/r^2$.
The disadvantage is that $\beta \sim 0.01$, which is outside the range given by Eq.~\eqref{range}. There is also an ambiguity in the force between particles with different mass,  this leading to a violation of Newton’s 3rd law.  In this model there should no GUP effect for neutral particles. \\

{\it Model 4.3}.  Associating the extra force with the strong interaction has the advantage that $\beta$ is in the range required by Eq.~\eqref{range}. It also links with the strong gravity proposal and has the interesting feature that  the GUP parameter depends on the type of particle and has a contribution from each parameter.  The disadvantage is that the extra force declines as $1/r^2$, rather than falling off exponentially beyond the hadron scale. One therefore needs to add some extra feature to avoid this.\\

{\it Model 4.4}.  Associating the extra force with the (classically well understood) irreducible mass $M_{irr}$, this just being the mass at infinity, has the advantage that there may be  no extra force at all.  Although the relationship between $M_{\rm ADM}$ and $M_{\rm irr}$ resembles the $M+1/M$ form, this seems to be a coincidence and one still needs the horizon radius to have the form $M_{irr} + 1/M_{irr}$. It is also interesting that $M=M_{irr}$ for the critical value of $\beta$ in CMMN analysis which separates two classes of solution. \\

{\it Model 4.5}.  Associating the extra force with another asymptotic space has the advantage that there is no extra force 
within {\it our} space.  It also  has the interesting feature that elementary particles are 
black holes.  The disadvantage is that this proposal (at least in its original form) is associated with LQG and a quadratic version of the GUP, with the status of black holes in this case being controversial.  However, this approach may also apply in a linear version of the GUP, where these problems do not arise.  So this may help model 4.4.\\
 
{\it Model 4.6}.  Associating the extra force with compactified higher dimensions has the advantage that this may give a cut-off in the force on the scale of the extra dimensions in the brane scenario. This may therefore resolve the problem associated with the second scenario. 
The disadvantage is that there is disagreement about the form of the GUP for higher dimensions.\\

We note that these different approaches all correspond to modifying the metric is some way.  In models 4.2 and 4.3, 
one puts $g_{00} = 1 - 2GM_{\rm ADM}/r$, so the horizon becomes $r_+ = 2GM_{\rm ADM}$, but there is an anomalous force between particles and the particles are associated with sub-Planckian black holes.  
In model 4.5, 
$M$ is unchanged but the circumferential radius (which depends on $g_{\theta \theta}$) becomes $R = (r^n + 1/r^n)^{1/n}$ where $n$ is 1 or 2. This implies that the event horizon radius is $r_+ = (M^n + 1/M^n)^{1/n}$ but there is no anomalous force between particles  in {\it our} space and particles are associated with
wormholes rather than black holes.  
In the 5-dimensional Kaluza-Klein approach of model 4.6,  $g_{rr}$ but not $g_{tt}$ is modified.
Of course,  the most satisfactory solution may involve a combination of these approaches and some of them are certainly connected.   For example,  one may combine the approaches in models 4.2 and 4.4 or in models 4.3 and 4.6.  

\section*{Acknowledgements}

The work of P.N. has partially
been supported by GNFM, Italy's National Group for Mathematical Physics.  JM is a KITP Scholar at the Kavli Institute for Theoretical Physics. The KITP Scholars Program is supported in part by the National Science Foundation under Grant No.~NSF PHY-1748958.
BC and JM thank the Frankfurt Institute for Advanced Studies, the Carl-Wilhelm-F\"uck Stiftung and the Department of Physics, University of Trieste for their generous hospitality, at which this work was partially written.

\bibliographystyle{apsrev4-1}
\bibliography{myrefs}

\begin{thebibliography}{80}%
\makeatletter
\providecommand \@ifxundefined [1]{%
 \@ifx{#1\undefined}
}%
\providecommand \@ifnum [1]{%
 \ifnum #1\expandafter \@firstoftwo
 \else \expandafter \@secondoftwo
 \fi
}%
\providecommand \@ifx [1]{%
 \ifx #1\expandafter \@firstoftwo
 \else \expandafter \@secondoftwo
 \fi
}%
\providecommand \natexlab [1]{#1}%
\providecommand \enquote  [1]{``#1''}%
\providecommand \bibnamefont  [1]{#1}%
\providecommand \bibfnamefont [1]{#1}%
\providecommand \citenamefont [1]{#1}%
\providecommand \href@noop [0]{\@secondoftwo}%
\providecommand \href [0]{\begingroup \@sanitize@url \@href}%
\providecommand \@href[1]{\@@startlink{#1}\@@href}%
\providecommand \@@href[1]{\endgroup#1\@@endlink}%
\providecommand \@sanitize@url [0]{\catcode `\\12\catcode `\$12\catcode
  `\&12\catcode `\#12\catcode `\^12\catcode `\_12\catcode `\%12\relax}%
\providecommand \@@startlink[1]{}%
\providecommand \@@endlink[0]{}%
\providecommand \url  [0]{\begingroup\@sanitize@url \@url }%
\providecommand \@url [1]{\endgroup\@href {#1}{\urlprefix }}%
\providecommand \urlprefix  [0]{URL }%
\providecommand \Eprint [0]{\href }%
\providecommand \doibase [0]{http://dx.doi.org/}%
\providecommand \selectlanguage [0]{\@gobble}%
\providecommand \bibinfo  [0]{\@secondoftwo}%
\providecommand \bibfield  [0]{\@secondoftwo}%
\providecommand \translation [1]{[#1]}%
\providecommand \BibitemOpen [0]{}%
\providecommand \bibitemStop [0]{}%
\providecommand \bibitemNoStop [0]{.\EOS\space}%
\providecommand \EOS [0]{\spacefactor3000\relax}%
\providecommand \BibitemShut  [1]{\csname bibitem#1\endcsname}%
\let\auto@bib@innerbib\@empty
%</preamble>
\bibitem [{\citenamefont {Garay}(1995)}]{Gar95}%
  \BibitemOpen
  \bibfield  {author} {\bibinfo {author} {\bibfnamefont {L.~J.}\ \bibnamefont
  {Garay}},\ }\href {\doibase 10.1142/S0217751X95000085} {\bibfield  {journal}
  {\bibinfo  {journal} {Int. J. Mod. Phys. A}\ }\textbf {\bibinfo {volume}
  {10}},\ \bibinfo {pages} {145} (\bibinfo {year} {1995})},\ \Eprint
  {http://arxiv.org/abs/gr-qc/9403008} {arXiv:gr-qc/9403008} \BibitemShut
  {NoStop}%
\bibitem [{\citenamefont {Padmanabhan}(1997)}]{Pad97}%
  \BibitemOpen
  \bibfield  {author} {\bibinfo {author} {\bibfnamefont {T.}~\bibnamefont
  {Padmanabhan}},\ }\href {\doibase 10.1103/PhysRevLett.78.1854} {\bibfield
  {journal} {\bibinfo  {journal} {Phys. Rev. Lett.}\ }\textbf {\bibinfo
  {volume} {78}},\ \bibinfo {pages} {1854} (\bibinfo {year} {1997})},\ \Eprint
  {http://arxiv.org/abs/hep-th/9608182} {arXiv:hep-th/9608182} \BibitemShut
  {NoStop}%
\bibitem [{\citenamefont {Nicolini}(2009)}]{Nic09}%
  \BibitemOpen
  \bibfield  {author} {\bibinfo {author} {\bibfnamefont {P.}~\bibnamefont
  {Nicolini}},\ }\href {\doibase 10.1142/S0217751X09043353} {\bibfield
  {journal} {\bibinfo  {journal} {Int. J. Mod. Phys.}\ }\textbf {\bibinfo
  {volume} {A24}},\ \bibinfo {pages} {1229} (\bibinfo {year} {2009})},\ \Eprint
  {http://arxiv.org/abs/0807.1939} {arXiv:0807.1939 [hep-th]} \BibitemShut
  {NoStop}%
%%CITATION = ARXIV:0807.1939;%%
\bibitem [{\citenamefont {Dvali}\ \emph {et~al.}(2011)\citenamefont {Dvali},
  \citenamefont {Folkerts},\ and\ \citenamefont {Germani}}]{DFG11}%
  \BibitemOpen
  \bibfield  {author} {\bibinfo {author} {\bibfnamefont {G.}~\bibnamefont
  {Dvali}}, \bibinfo {author} {\bibfnamefont {S.}~\bibnamefont {Folkerts}}, \
  and\ \bibinfo {author} {\bibfnamefont {C.}~\bibnamefont {Germani}},\ }\href
  {\doibase 10.1103/PhysRevD.84.024039} {\bibfield  {journal} {\bibinfo
  {journal} {Phys. Rev. D}\ }\textbf {\bibinfo {volume} {84}},\ \bibinfo
  {pages} {024039} (\bibinfo {year} {2011})},\ \Eprint
  {http://arxiv.org/abs/1006.0984} {arXiv:1006.0984 [hep-th]} \BibitemShut
  {NoStop}%
\bibitem [{\citenamefont {Rovelli}(1998)}]{Rov98}%
  \BibitemOpen
  \bibfield  {author} {\bibinfo {author} {\bibfnamefont {C.}~\bibnamefont
  {Rovelli}},\ }\href {\doibase 10.12942/lrr-1998-1} {\bibfield  {journal}
  {\bibinfo  {journal} {Living Rev. Rel.}\ }\textbf {\bibinfo {volume} {1}},\
  \bibinfo {pages} {1} (\bibinfo {year} {1998})},\ \Eprint
  {http://arxiv.org/abs/gr-qc/9710008} {arXiv:gr-qc/9710008} \BibitemShut
  {NoStop}%
\bibitem [{\citenamefont {Sprenger}\ \emph {et~al.}(2012)\citenamefont
  {Sprenger}, \citenamefont {Nicolini},\ and\ \citenamefont
  {Bleicher}}]{SNB12}%
  \BibitemOpen
  \bibfield  {author} {\bibinfo {author} {\bibfnamefont {M.}~\bibnamefont
  {Sprenger}}, \bibinfo {author} {\bibfnamefont {P.}~\bibnamefont {Nicolini}},
  \ and\ \bibinfo {author} {\bibfnamefont {M.}~\bibnamefont {Bleicher}},\
  }\href {\doibase 10.1088/0143-0807/33/4/853} {\bibfield  {journal} {\bibinfo
  {journal} {Eur. J. Phys.}\ }\textbf {\bibinfo {volume} {33}},\ \bibinfo
  {pages} {853} (\bibinfo {year} {2012})},\ \Eprint
  {http://arxiv.org/abs/1202.1500} {arXiv:1202.1500 [physics.ed-ph]}
  \BibitemShut {NoStop}%
\bibitem [{\citenamefont {Hossenfelder}(2013)}]{Hos13}%
  \BibitemOpen
  \bibfield  {author} {\bibinfo {author} {\bibfnamefont {S.}~\bibnamefont
  {Hossenfelder}},\ }\href {\doibase 10.12942/lrr-2013-2} {\bibfield  {journal}
  {\bibinfo  {journal} {Living Rev. Rel.}\ }\textbf {\bibinfo {volume} {16}},\
  \bibinfo {pages} {2} (\bibinfo {year} {2013})},\ \Eprint
  {http://arxiv.org/abs/1203.6191} {arXiv:1203.6191 [gr-qc]} \BibitemShut
  {NoStop}%
\bibitem [{\citenamefont {Amati}\ \emph {et~al.}(1987)\citenamefont {Amati},
  \citenamefont {Ciafaloni},\ and\ \citenamefont {Veneziano}}]{ACV87}%
  \BibitemOpen
  \bibfield  {author} {\bibinfo {author} {\bibfnamefont {D.}~\bibnamefont
  {Amati}}, \bibinfo {author} {\bibfnamefont {M.}~\bibnamefont {Ciafaloni}}, \
  and\ \bibinfo {author} {\bibfnamefont {G.}~\bibnamefont {Veneziano}},\ }\href
  {\doibase 10.1016/0370-2693(87)90346-7} {\bibfield  {journal} {\bibinfo
  {journal} {Phys. Lett. B}\ }\textbf {\bibinfo {volume} {197}},\ \bibinfo
  {pages} {81} (\bibinfo {year} {1987})}\BibitemShut {NoStop}%
\bibitem [{\citenamefont {Amati}\ \emph {et~al.}(1988)\citenamefont {Amati},
  \citenamefont {Ciafaloni},\ and\ \citenamefont {Veneziano}}]{ACV88}%
  \BibitemOpen
  \bibfield  {author} {\bibinfo {author} {\bibfnamefont {D.}~\bibnamefont
  {Amati}}, \bibinfo {author} {\bibfnamefont {M.}~\bibnamefont {Ciafaloni}}, \
  and\ \bibinfo {author} {\bibfnamefont {G.}~\bibnamefont {Veneziano}},\ }\href
  {\doibase 10.1142/S0217751X88000710} {\bibfield  {journal} {\bibinfo
  {journal} {Int. J. Mod. Phys. A}\ }\textbf {\bibinfo {volume} {3}},\ \bibinfo
  {pages} {1615} (\bibinfo {year} {1988})}\BibitemShut {NoStop}%
\bibitem [{\citenamefont {Amati}\ \emph {et~al.}(1989)\citenamefont {Amati},
  \citenamefont {Ciafaloni},\ and\ \citenamefont {Veneziano}}]{ACV89}%
  \BibitemOpen
  \bibfield  {author} {\bibinfo {author} {\bibfnamefont {D.}~\bibnamefont
  {Amati}}, \bibinfo {author} {\bibfnamefont {M.}~\bibnamefont {Ciafaloni}}, \
  and\ \bibinfo {author} {\bibfnamefont {G.}~\bibnamefont {Veneziano}},\ }\href
  {\doibase 10.1016/0370-2693(89)91366-X} {\bibfield  {journal} {\bibinfo
  {journal} {Phys. Lett. B}\ }\textbf {\bibinfo {volume} {216}},\ \bibinfo
  {pages} {41} (\bibinfo {year} {1989})}\BibitemShut {NoStop}%
\bibitem [{\citenamefont {Maggiore}(1993)}]{Mag93}%
  \BibitemOpen
  \bibfield  {author} {\bibinfo {author} {\bibfnamefont {M.}~\bibnamefont
  {Maggiore}},\ }\href {\doibase 10.1016/0370-2693(93)91401-8} {\bibfield
  {journal} {\bibinfo  {journal} {Phys. Lett. B}\ }\textbf {\bibinfo {volume}
  {304}},\ \bibinfo {pages} {65} (\bibinfo {year} {1993})},\ \Eprint
  {http://arxiv.org/abs/hep-th/9301067} {arXiv:hep-th/9301067} \BibitemShut
  {NoStop}%
\bibitem [{\citenamefont {Kempf}\ \emph {et~al.}(1995)\citenamefont {Kempf},
  \citenamefont {Mangano},\ and\ \citenamefont {Mann}}]{KMM95}%
  \BibitemOpen
  \bibfield  {author} {\bibinfo {author} {\bibfnamefont {A.}~\bibnamefont
  {Kempf}}, \bibinfo {author} {\bibfnamefont {G.}~\bibnamefont {Mangano}}, \
  and\ \bibinfo {author} {\bibfnamefont {R.~B.}\ \bibnamefont {Mann}},\ }\href
  {\doibase 10.1103/PhysRevD.52.1108} {\bibfield  {journal} {\bibinfo
  {journal} {Phys. Rev. D}\ }\textbf {\bibinfo {volume} {52}},\ \bibinfo
  {pages} {1108} (\bibinfo {year} {1995})},\ \Eprint
  {http://arxiv.org/abs/hep-th/9412167} {arXiv:hep-th/9412167} \BibitemShut
  {NoStop}%
\bibitem [{\citenamefont {Adler}\ and\ \citenamefont {Santiago}(1999)}]{AdS99}%
  \BibitemOpen
  \bibfield  {author} {\bibinfo {author} {\bibfnamefont {R.~J.}\ \bibnamefont
  {Adler}}\ and\ \bibinfo {author} {\bibfnamefont {D.~I.}\ \bibnamefont
  {Santiago}},\ }\href {\doibase 10.1142/S0217732399001462} {\bibfield
  {journal} {\bibinfo  {journal} {Mod. Phys. Lett. A}\ }\textbf {\bibinfo
  {volume} {14}},\ \bibinfo {pages} {1371} (\bibinfo {year} {1999})},\ \Eprint
  {http://arxiv.org/abs/gr-qc/9904026} {arXiv:gr-qc/9904026} \BibitemShut
  {NoStop}%
\bibitem [{\citenamefont {Adler}\ \emph {et~al.}(2001)\citenamefont {Adler},
  \citenamefont {Chen},\ and\ \citenamefont {Santiago}}]{ACS01}%
  \BibitemOpen
  \bibfield  {author} {\bibinfo {author} {\bibfnamefont {R.~J.}\ \bibnamefont
  {Adler}}, \bibinfo {author} {\bibfnamefont {P.}~\bibnamefont {Chen}}, \ and\
  \bibinfo {author} {\bibfnamefont {D.~I.}\ \bibnamefont {Santiago}},\ }\href
  {\doibase 10.1023/A:1015281430411} {\bibfield  {journal} {\bibinfo  {journal}
  {Gen. Rel. Grav.}\ }\textbf {\bibinfo {volume} {33}},\ \bibinfo {pages}
  {2101} (\bibinfo {year} {2001})},\ \Eprint
  {http://arxiv.org/abs/gr-qc/0106080} {arXiv:gr-qc/0106080} \BibitemShut
  {NoStop}%
\bibitem [{\citenamefont {Chen}\ and\ \citenamefont {Adler}(2003)}]{ChA03}%
  \BibitemOpen
  \bibfield  {author} {\bibinfo {author} {\bibfnamefont {P.}~\bibnamefont
  {Chen}}\ and\ \bibinfo {author} {\bibfnamefont {R.~J.}\ \bibnamefont
  {Adler}},\ }\href {\doibase 10.1016/S0920-5632(03)02088-7} {\bibfield
  {journal} {\bibinfo  {journal} {Nucl. Phys. B Proc. Suppl.}\ }\textbf
  {\bibinfo {volume} {124}},\ \bibinfo {pages} {103} (\bibinfo {year}
  {2003})},\ \Eprint {http://arxiv.org/abs/gr-qc/0205106} {arXiv:gr-qc/0205106}
  \BibitemShut {NoStop}%
\bibitem [{\citenamefont {Adler}(2010)}]{Adl10}%
  \BibitemOpen
  \bibfield  {author} {\bibinfo {author} {\bibfnamefont {R.~J.}\ \bibnamefont
  {Adler}},\ }\href {\doibase 10.1119/1.3439650} {\bibfield  {journal}
  {\bibinfo  {journal} {Am. J. Phys.}\ }\textbf {\bibinfo {volume} {78}},\
  \bibinfo {pages} {925} (\bibinfo {year} {2010})},\ \Eprint
  {http://arxiv.org/abs/1001.1205} {arXiv:1001.1205 [gr-qc]} \BibitemShut
  {NoStop}%
\bibitem [{\citenamefont {Isi}\ \emph {et~al.}(2013)\citenamefont {Isi},
  \citenamefont {Mureika},\ and\ \citenamefont {Nicolini}}]{IMN13}%
  \BibitemOpen
  \bibfield  {author} {\bibinfo {author} {\bibfnamefont {M.}~\bibnamefont
  {Isi}}, \bibinfo {author} {\bibfnamefont {J.}~\bibnamefont {Mureika}}, \ and\
  \bibinfo {author} {\bibfnamefont {P.}~\bibnamefont {Nicolini}},\ }\href
  {\doibase 10.1007/JHEP11(2013)139} {\bibfield  {journal} {\bibinfo  {journal}
  {JHEP}\ }\textbf {\bibinfo {volume} {11}},\ \bibinfo {pages} {139} (\bibinfo
  {year} {2013})},\ \Eprint {http://arxiv.org/abs/1310.8153} {arXiv:1310.8153
  [hep-th]} \BibitemShut {NoStop}%
%%CITATION = ARXIV:1310.8153;%%
\bibitem [{\citenamefont {Carr}\ \emph {et~al.}(2011)\citenamefont {Carr},
  \citenamefont {Modesto},\ and\ \citenamefont {Premont-Schwarz}}]{CMP11}%
  \BibitemOpen
  \bibfield  {author} {\bibinfo {author} {\bibfnamefont {B.}~\bibnamefont
  {Carr}}, \bibinfo {author} {\bibfnamefont {L.}~\bibnamefont {Modesto}}, \
  and\ \bibinfo {author} {\bibfnamefont {I.}~\bibnamefont {Premont-Schwarz}},\
  }\href@noop {} {\  (\bibinfo {year} {2011})},\ \Eprint
  {http://arxiv.org/abs/1107.0708} {arXiv:1107.0708 [gr-qc]} \BibitemShut
  {NoStop}%
\bibitem [{\citenamefont {Carr}(2016)}]{Carr16}%
  \BibitemOpen
  \bibfield  {author} {\bibinfo {author} {\bibfnamefont {B.~J.}\ \bibnamefont
  {Carr}},\ }\href {\doibase 10.1007/978-3-319-20046-0_19} {\bibfield
  {journal} {\bibinfo  {journal} {Springer Proc. Phys.}\ }\textbf {\bibinfo
  {volume} {170}},\ \bibinfo {pages} {159} (\bibinfo {year} {2016})},\ \Eprint
  {http://arxiv.org/abs/1402.1427} {arXiv:1402.1427 [gr-qc]} \BibitemShut
  {NoStop}%
\bibitem [{\citenamefont {Carr}\ \emph {et~al.}(2015)\citenamefont {Carr},
  \citenamefont {Mureika},\ and\ \citenamefont {Nicolini}}]{CMN15}%
  \BibitemOpen
  \bibfield  {author} {\bibinfo {author} {\bibfnamefont {B.~J.}\ \bibnamefont
  {Carr}}, \bibinfo {author} {\bibfnamefont {J.}~\bibnamefont {Mureika}}, \
  and\ \bibinfo {author} {\bibfnamefont {P.}~\bibnamefont {Nicolini}},\ }\href
  {\doibase 10.1007/JHEP07(2015)052} {\bibfield  {journal} {\bibinfo  {journal}
  {JHEP}\ }\textbf {\bibinfo {volume} {07}},\ \bibinfo {pages} {052} (\bibinfo
  {year} {2015})},\ \Eprint {http://arxiv.org/abs/1504.07637} {arXiv:1504.07637
  [gr-qc]} \BibitemShut {NoStop}%
%%CITATION = ARXIV:1504.07637;%%
\bibitem [{\citenamefont {Carr}\ \emph {et~al.}(2020)\citenamefont {Carr},
  \citenamefont {Mentzer}, \citenamefont {Mureika},\ and\ \citenamefont
  {Nicolini}}]{CMMN20}%
  \BibitemOpen
  \bibfield  {author} {\bibinfo {author} {\bibfnamefont {B.}~\bibnamefont
  {Carr}}, \bibinfo {author} {\bibfnamefont {H.}~\bibnamefont {Mentzer}},
  \bibinfo {author} {\bibfnamefont {J.}~\bibnamefont {Mureika}}, \ and\
  \bibinfo {author} {\bibfnamefont {P.}~\bibnamefont {Nicolini}},\ }\href
  {\doibase 10.1140/epjc/s10052-020-08706-0} {\bibfield  {journal} {\bibinfo
  {journal} {Eur. Phys. J. C}\ }\textbf {\bibinfo {volume} {80}},\ \bibinfo
  {pages} {1166} (\bibinfo {year} {2020})},\ \Eprint
  {http://arxiv.org/abs/2006.04892} {arXiv:2006.04892 [gr-qc]} \BibitemShut
  {NoStop}%
\bibitem [{\citenamefont {Lake}\ and\ \citenamefont
  {Carr}(2015)}]{Lake:2015pma}%
  \BibitemOpen
  \bibfield  {author} {\bibinfo {author} {\bibfnamefont {M.~J.}\ \bibnamefont
  {Lake}}\ and\ \bibinfo {author} {\bibfnamefont {B.}~\bibnamefont {Carr}},\
  }\href {\doibase 10.1007/JHEP11(2015)105} {\bibfield  {journal} {\bibinfo
  {journal} {JHEP}\ }\textbf {\bibinfo {volume} {11}},\ \bibinfo {pages} {105}
  (\bibinfo {year} {2015})},\ \Eprint {http://arxiv.org/abs/1505.06994}
  {arXiv:1505.06994 [gr-qc]} \BibitemShut {NoStop}%
\bibitem [{\citenamefont {Lake}\ and\ \citenamefont {Carr}(2018)}]{LaC18}%
  \BibitemOpen
  \bibfield  {author} {\bibinfo {author} {\bibfnamefont {M.~J.}\ \bibnamefont
  {Lake}}\ and\ \bibinfo {author} {\bibfnamefont {B.}~\bibnamefont {Carr}},\
  }\href {\doibase 10.1142/S0218271819300015} {\bibfield  {journal} {\bibinfo
  {journal} {Int. J. Mod. Phys. D}\ }\textbf {\bibinfo {volume} {27}},\
  \bibinfo {pages} {1930001} (\bibinfo {year} {2018})},\ \Eprint
  {http://arxiv.org/abs/1808.08386} {arXiv:1808.08386 [gr-qc]} \BibitemShut
  {NoStop}%
\bibitem [{\citenamefont {Dvali}\ and\ \citenamefont {Gomez}(2012)}]{DvG12}%
  \BibitemOpen
  \bibfield  {author} {\bibinfo {author} {\bibfnamefont {G.}~\bibnamefont
  {Dvali}}\ and\ \bibinfo {author} {\bibfnamefont {C.}~\bibnamefont {Gomez}},\
  }\href@noop {} {\enquote {\bibinfo {title} {{Black Hole
  Macro-Quantumness}},}\ } (\bibinfo {year} {2012}),\ \Eprint
  {http://arxiv.org/abs/1212.0765} {arXiv:1212.0765 [hep-th]} \BibitemShut
  {NoStop}%
\bibitem [{\citenamefont {Dvali}\ and\ \citenamefont
  {Gomez}(2013{\natexlab{a}})}]{DvG13}%
  \BibitemOpen
  \bibfield  {author} {\bibinfo {author} {\bibfnamefont {G.}~\bibnamefont
  {Dvali}}\ and\ \bibinfo {author} {\bibfnamefont {C.}~\bibnamefont {Gomez}},\
  }\href {\doibase 10.1002/prop.201300001} {\bibfield  {journal} {\bibinfo
  {journal} {Fortsch. Phys.}\ }\textbf {\bibinfo {volume} {61}},\ \bibinfo
  {pages} {742} (\bibinfo {year} {2013}{\natexlab{a}})},\ \Eprint
  {http://arxiv.org/abs/1112.3359} {arXiv:1112.3359 [hep-th]} \BibitemShut
  {NoStop}%
\bibitem [{\citenamefont {Dvali}\ and\ \citenamefont
  {Gomez}(2013{\natexlab{b}})}]{DvG13+}%
  \BibitemOpen
  \bibfield  {author} {\bibinfo {author} {\bibfnamefont {G.}~\bibnamefont
  {Dvali}}\ and\ \bibinfo {author} {\bibfnamefont {C.}~\bibnamefont {Gomez}},\
  }\href {\doibase 10.1016/j.physletb.2013.01.020} {\bibfield  {journal}
  {\bibinfo  {journal} {Phys. Lett.}\ }\textbf {\bibinfo {volume} {B719}},\
  \bibinfo {pages} {419} (\bibinfo {year} {2013}{\natexlab{b}})},\ \Eprint
  {http://arxiv.org/abs/1203.6575} {arXiv:1203.6575 [hep-th]} \BibitemShut
  {NoStop}%
%%CITATION = ARXIV:1203.6575;%%
\bibitem [{\citenamefont {Nicolini}\ and\ \citenamefont
  {Spallucci}(2014)}]{NiS14}%
  \BibitemOpen
  \bibfield  {author} {\bibinfo {author} {\bibfnamefont {P.}~\bibnamefont
  {Nicolini}}\ and\ \bibinfo {author} {\bibfnamefont {E.}~\bibnamefont
  {Spallucci}},\ }\href {\doibase 10.1155/2014/805684} {\bibfield  {journal}
  {\bibinfo  {journal} {Adv. High Energy Phys.}\ }\textbf {\bibinfo {volume}
  {2014}},\ \bibinfo {pages} {805684} (\bibinfo {year} {2014})},\ \Eprint
  {http://arxiv.org/abs/1210.0015} {arXiv:1210.0015 [hep-th]} \BibitemShut
  {NoStop}%
%%CITATION = ARXIV:1210.0015;%%
\bibitem [{\citenamefont {Nicolini}(2023)}]{Nicolini:2023hub}%
  \BibitemOpen
  \bibfield  {author} {\bibinfo {author} {\bibfnamefont {P.}~\bibnamefont
  {Nicolini}},\ }\enquote {\bibinfo {title} {{How strings can explain regular
  black holes}},}\ in\ \href {\doibase 10.1007/978-981-99-1596-5_3} {\emph
  {\bibinfo {booktitle} {Regular Black Holes: Towards a New Paradigm of
  Gravitational Collapse}}},\ \bibinfo {editor} {edited by\ \bibinfo {editor}
  {\bibfnamefont {C.}~\bibnamefont {Bambi}}}\ (\bibinfo  {publisher} {Springer,
  Singapore},\ \bibinfo {year} {2023})\ Chap.~\bibinfo {chapter} {{}},\ \Eprint
  {http://arxiv.org/abs/2306.01480} {arXiv:2306.01480 [gr-qc]} \BibitemShut
  {NoStop}%
\bibitem [{\citenamefont {'t~Hooft}\ \emph {et~al.}(2018)\citenamefont
  {'t~Hooft}, \citenamefont {Giddings}, \citenamefont {Rovelli}, \citenamefont
  {Nicolini}, \citenamefont {Mureika}, \citenamefont {Kaminski},\ and\
  \citenamefont {Bleicher}}]{tGR+18}%
  \BibitemOpen
  \bibfield  {author} {\bibinfo {author} {\bibfnamefont {G.}~\bibnamefont
  {'t~Hooft}}, \bibinfo {author} {\bibfnamefont {S.~B.}\ \bibnamefont
  {Giddings}}, \bibinfo {author} {\bibfnamefont {C.}~\bibnamefont {Rovelli}},
  \bibinfo {author} {\bibfnamefont {P.}~\bibnamefont {Nicolini}}, \bibinfo
  {author} {\bibfnamefont {J.}~\bibnamefont {Mureika}}, \bibinfo {author}
  {\bibfnamefont {M.}~\bibnamefont {Kaminski}}, \ and\ \bibinfo {author}
  {\bibfnamefont {M.}~\bibnamefont {Bleicher}},\ }\enquote {\bibinfo {title}
  {{Panel Discussion, ``The Duel'': The Good, the Bad, and the Ugly of Gravity
  and Information}},}\ in\ \href {\doibase 10.1007/978-3-319-94256-8\_2} {\emph
  {\bibinfo {booktitle} {2nd Karl Schwarzschild Meeting on Gravitational
  Physics}}},\ \bibinfo {series} {Springer Proc. Phys.}, Vol.\ \bibinfo
  {volume} {208},\ \bibinfo {editor} {edited by\ \bibinfo {editor}
  {\bibfnamefont {P.}~\bibnamefont {Nicolini}}, \bibinfo {editor}
  {\bibfnamefont {M.}~\bibnamefont {Kaminski}}, \bibinfo {editor}
  {\bibfnamefont {J.}~\bibnamefont {Mureika}}, \ and\ \bibinfo {editor}
  {\bibfnamefont {M.}~\bibnamefont {Bleicher}}}\ (\bibinfo  {publisher}
  {Springer},\ \bibinfo {year} {2018})\ pp.\ \bibinfo {pages} {13--35},\
  \Eprint {http://arxiv.org/abs/1609.01725} {arXiv:1609.01725 [hep-th]}
  \BibitemShut {NoStop}%
\bibitem [{\citenamefont {Giddings}(2017)}]{Gid17}%
  \BibitemOpen
  \bibfield  {author} {\bibinfo {author} {\bibfnamefont {S.~B.}\ \bibnamefont
  {Giddings}},\ }\href {\doibase 10.1038/s41550-017-0067} {\bibfield  {journal}
  {\bibinfo  {journal} {Nature Astron.}\ }\textbf {\bibinfo {volume} {1}},\
  \bibinfo {pages} {0067} (\bibinfo {year} {2017})},\ \Eprint
  {http://arxiv.org/abs/1703.03387} {arXiv:1703.03387 [gr-qc]} \BibitemShut
  {NoStop}%
\bibitem [{\citenamefont {Bonanno}\ and\ \citenamefont {Reuter}(2006)}]{BoR06}%
  \BibitemOpen
  \bibfield  {author} {\bibinfo {author} {\bibfnamefont {A.}~\bibnamefont
  {Bonanno}}\ and\ \bibinfo {author} {\bibfnamefont {M.}~\bibnamefont
  {Reuter}},\ }\href {\doibase 10.1103/PhysRevD.73.083005} {\bibfield
  {journal} {\bibinfo  {journal} {Phys. Rev. D}\ }\textbf {\bibinfo {volume}
  {73}},\ \bibinfo {pages} {083005} (\bibinfo {year} {2006})},\ \Eprint
  {http://arxiv.org/abs/hep-th/0602159} {arXiv:hep-th/0602159} \BibitemShut
  {NoStop}%
\bibitem [{\citenamefont {Frassino}\ \emph {et~al.}(2016)\citenamefont
  {Frassino}, \citenamefont {Koeppel},\ and\ \citenamefont {Nicolini}}]{FKN16}%
  \BibitemOpen
  \bibfield  {author} {\bibinfo {author} {\bibfnamefont {A.~M.}\ \bibnamefont
  {Frassino}}, \bibinfo {author} {\bibfnamefont {S.}~\bibnamefont {Koeppel}}, \
  and\ \bibinfo {author} {\bibfnamefont {P.}~\bibnamefont {Nicolini}},\ }\href
  {\doibase 10.3390/e18050181} {\bibfield  {journal} {\bibinfo  {journal}
  {Entropy}\ }\textbf {\bibinfo {volume} {18}},\ \bibinfo {pages} {181}
  (\bibinfo {year} {2016})},\ \Eprint {http://arxiv.org/abs/1604.03263}
  {arXiv:1604.03263 [gr-qc]} \BibitemShut {NoStop}%
%%CITATION = ARXIV:1604.03263;%%
\bibitem [{\citenamefont {Spallucci}\ and\ \citenamefont
  {Smailagic}(2021)}]{SpS21}%
  \BibitemOpen
  \bibfield  {author} {\bibinfo {author} {\bibfnamefont {E.}~\bibnamefont
  {Spallucci}}\ and\ \bibinfo {author} {\bibfnamefont {A.}~\bibnamefont
  {Smailagic}},\ }\href {\doibase 10.1016/j.physletb.2021.136180} {\bibfield
  {journal} {\bibinfo  {journal} {Phys. Lett. B}\ }\textbf {\bibinfo {volume}
  {816}},\ \bibinfo {pages} {136180} (\bibinfo {year} {2021})},\ \Eprint
  {http://arxiv.org/abs/2103.03947} {arXiv:2103.03947 [hep-th]} \BibitemShut
  {NoStop}%
\bibitem [{\citenamefont {Carr}(2022)}]{Carr:2022ndy}%
  \BibitemOpen
  \bibfield  {author} {\bibinfo {author} {\bibfnamefont {B.~J.}\ \bibnamefont
  {Carr}},\ }\href {\doibase 10.3389/fspas.2022.1008221} {\bibfield  {journal}
  {\bibinfo  {journal} {Front. Astron. Space Sci.}\ }\textbf {\bibinfo {volume}
  {9}},\ \bibinfo {pages} {1008221} (\bibinfo {year} {2022})},\ \Eprint
  {http://arxiv.org/abs/2302.12609} {arXiv:2302.12609 [gr-qc]} \BibitemShut
  {NoStop}%
\bibitem [{\citenamefont {Mureika}\ and\ \citenamefont
  {Nicolini}(2013)}]{MuN13}%
  \BibitemOpen
  \bibfield  {author} {\bibinfo {author} {\bibfnamefont {J.}~\bibnamefont
  {Mureika}}\ and\ \bibinfo {author} {\bibfnamefont {P.}~\bibnamefont
  {Nicolini}},\ }\href {\doibase 10.1140/epjp/i2013-13078-0} {\bibfield
  {journal} {\bibinfo  {journal} {Eur. Phys. J. Plus}\ }\textbf {\bibinfo
  {volume} {128}},\ \bibinfo {pages} {78} (\bibinfo {year} {2013})},\ \Eprint
  {http://arxiv.org/abs/1206.4696} {arXiv:1206.4696 [hep-th]} \BibitemShut
  {NoStop}%
%%CITATION = ARXIV:1206.4696;%%
\bibitem [{\citenamefont {Buoninfante}\ \emph {et~al.}(2019)\citenamefont
  {Buoninfante}, \citenamefont {Luciano},\ and\ \citenamefont
  {Petruzziello}}]{Buoninfante_2019}%
  \BibitemOpen
  \bibfield  {author} {\bibinfo {author} {\bibfnamefont {L.}~\bibnamefont
  {Buoninfante}}, \bibinfo {author} {\bibfnamefont {G.~G.}\ \bibnamefont
  {Luciano}}, \ and\ \bibinfo {author} {\bibfnamefont {L.}~\bibnamefont
  {Petruzziello}},\ }\href {\doibase 10.1140/epjc/s10052-019-7164-y} {\bibfield
   {journal} {\bibinfo  {journal} {The European Physical Journal C}\ }\textbf
  {\bibinfo {volume} {79}} (\bibinfo {year} {2019}),\
  10.1140/epjc/s10052-019-7164-y}\BibitemShut {NoStop}%
\bibitem [{\citenamefont {Bishop}\ \emph {et~al.}(2023)\citenamefont {Bishop},
  \citenamefont {Contreras}, \citenamefont {Martin}, \citenamefont {Nicolini},\
  and\ \citenamefont {Singleton}}]{Bishop:2023hvz}%
  \BibitemOpen
  \bibfield  {author} {\bibinfo {author} {\bibfnamefont {M.}~\bibnamefont
  {Bishop}}, \bibinfo {author} {\bibfnamefont {J.}~\bibnamefont {Contreras}},
  \bibinfo {author} {\bibfnamefont {P.}~\bibnamefont {Martin}}, \bibinfo
  {author} {\bibfnamefont {P.}~\bibnamefont {Nicolini}}, \ and\ \bibinfo
  {author} {\bibfnamefont {D.}~\bibnamefont {Singleton}},\ }\href {\doibase
  10.1016/j.physletb.2023.138263} {\bibfield  {journal} {\bibinfo  {journal}
  {Phys. Lett. B}\ }\textbf {\bibinfo {volume} {847}},\ \bibinfo {pages}
  {138263} (\bibinfo {year} {2023})},\ \Eprint
  {http://arxiv.org/abs/2307.05367} {arXiv:2307.05367 [quant-ph]} \BibitemShut
  {NoStop}%
\bibitem [{\citenamefont {Das}\ and\ \citenamefont
  {Vagenas}(2008)}]{Das:2008kaa}%
  \BibitemOpen
  \bibfield  {author} {\bibinfo {author} {\bibfnamefont {S.}~\bibnamefont
  {Das}}\ and\ \bibinfo {author} {\bibfnamefont {E.~C.}\ \bibnamefont
  {Vagenas}},\ }\href {\doibase 10.1103/PhysRevLett.101.221301} {\bibfield
  {journal} {\bibinfo  {journal} {Phys. Rev. Lett.}\ }\textbf {\bibinfo
  {volume} {101}},\ \bibinfo {pages} {221301} (\bibinfo {year} {2008})},\
  \Eprint {http://arxiv.org/abs/0810.5333} {arXiv:0810.5333 [hep-th]}
  \BibitemShut {NoStop}%
\bibitem [{\citenamefont {Pikovski}\ \emph {et~al.}(2012)\citenamefont
  {Pikovski}, \citenamefont {Vanner}, \citenamefont {Aspelmeyer}, \citenamefont
  {Kim},\ and\ \citenamefont {Brukner}}]{Pikovski:2011zk}%
  \BibitemOpen
  \bibfield  {author} {\bibinfo {author} {\bibfnamefont {I.}~\bibnamefont
  {Pikovski}}, \bibinfo {author} {\bibfnamefont {M.~R.}\ \bibnamefont
  {Vanner}}, \bibinfo {author} {\bibfnamefont {M.}~\bibnamefont {Aspelmeyer}},
  \bibinfo {author} {\bibfnamefont {M.}~\bibnamefont {Kim}}, \ and\ \bibinfo
  {author} {\bibfnamefont {C.}~\bibnamefont {Brukner}},\ }\href {\doibase
  10.1038/nphys2262} {\bibfield  {journal} {\bibinfo  {journal} {Nature Phys.}\
  }\textbf {\bibinfo {volume} {8}},\ \bibinfo {pages} {393} (\bibinfo {year}
  {2012})},\ \Eprint {http://arxiv.org/abs/1111.1979} {arXiv:1111.1979
  [quant-ph]} \BibitemShut {NoStop}%
\bibitem [{\citenamefont {Kumar}\ and\ \citenamefont
  {Plenio}(2020)}]{Kumar:2019bnd}%
  \BibitemOpen
  \bibfield  {author} {\bibinfo {author} {\bibfnamefont {S.~P.}\ \bibnamefont
  {Kumar}}\ and\ \bibinfo {author} {\bibfnamefont {M.~B.}\ \bibnamefont
  {Plenio}},\ }\href {\doibase 10.1038/s41467-020-17518-5} {\bibfield
  {journal} {\bibinfo  {journal} {Nature Commun.}\ }\textbf {\bibinfo {volume}
  {11}},\ \bibinfo {pages} {3900} (\bibinfo {year} {2020})},\ \Eprint
  {http://arxiv.org/abs/1908.11164} {arXiv:1908.11164 [quant-ph]} \BibitemShut
  {NoStop}%
\bibitem [{\citenamefont {Iorio}\ \emph {et~al.}(2018)\citenamefont {Iorio},
  \citenamefont {Pais}, \citenamefont {Elmashad}, \citenamefont {Ali},
  \citenamefont {Faizal},\ and\ \citenamefont {Abou-Salem}}]{Iorio:2017vtw}%
  \BibitemOpen
  \bibfield  {author} {\bibinfo {author} {\bibfnamefont {A.}~\bibnamefont
  {Iorio}}, \bibinfo {author} {\bibfnamefont {P.}~\bibnamefont {Pais}},
  \bibinfo {author} {\bibfnamefont {I.~A.}\ \bibnamefont {Elmashad}}, \bibinfo
  {author} {\bibfnamefont {A.~F.}\ \bibnamefont {Ali}}, \bibinfo {author}
  {\bibfnamefont {M.}~\bibnamefont {Faizal}}, \ and\ \bibinfo {author}
  {\bibfnamefont {L.~I.}\ \bibnamefont {Abou-Salem}},\ }\href {\doibase
  10.1142/S0218271818500803} {\bibfield  {journal} {\bibinfo  {journal} {Int.
  J. Mod. Phys. D}\ }\textbf {\bibinfo {volume} {27}},\ \bibinfo {pages}
  {1850080} (\bibinfo {year} {2018})},\ \bibinfo {note} {[Erratum:
  Int.J.Mod.Phys.D 27, 1850080 (2023)]},\ \Eprint
  {http://arxiv.org/abs/1706.01332} {arXiv:1706.01332 [physics.gen-ph]}
  \BibitemShut {NoStop}%
\bibitem [{\citenamefont {Bushev}\ \emph {et~al.}(2019)\citenamefont {Bushev},
  \citenamefont {Bourhill}, \citenamefont {Goryachev}, \citenamefont
  {Kukharchyk}, \citenamefont {Ivanov}, \citenamefont {Galliou}, \citenamefont
  {Tobar},\ and\ \citenamefont {Danilishin}}]{bushev19}%
  \BibitemOpen
  \bibfield  {author} {\bibinfo {author} {\bibfnamefont {P.~A.}\ \bibnamefont
  {Bushev}}, \bibinfo {author} {\bibfnamefont {J.}~\bibnamefont {Bourhill}},
  \bibinfo {author} {\bibfnamefont {M.}~\bibnamefont {Goryachev}}, \bibinfo
  {author} {\bibfnamefont {N.}~\bibnamefont {Kukharchyk}}, \bibinfo {author}
  {\bibfnamefont {E.}~\bibnamefont {Ivanov}}, \bibinfo {author} {\bibfnamefont
  {S.}~\bibnamefont {Galliou}}, \bibinfo {author} {\bibfnamefont {M.~E.}\
  \bibnamefont {Tobar}}, \ and\ \bibinfo {author} {\bibfnamefont
  {S.}~\bibnamefont {Danilishin}},\ }\href {\doibase
  10.1103/PhysRevD.100.066020} {\bibfield  {journal} {\bibinfo  {journal}
  {Phys. Rev. D}\ }\textbf {\bibinfo {volume} {100}},\ \bibinfo {pages}
  {066020} (\bibinfo {year} {2019})},\ \Eprint
  {http://arxiv.org/abs/1903.03346} {arXiv:1903.03346 [quant-ph]} \BibitemShut
  {NoStop}%
\bibitem [{\citenamefont {Bawaj}\ \emph {et~al.}(2015)\citenamefont {Bawaj}
  \emph {et~al.}}]{Bawaj:2014cda}%
  \BibitemOpen
  \bibfield  {author} {\bibinfo {author} {\bibfnamefont {M.}~\bibnamefont
  {Bawaj}} \emph {et~al.},\ }\href {\doibase 10.1038/ncomms8503} {\bibfield
  {journal} {\bibinfo  {journal} {Nature Commun.}\ }\textbf {\bibinfo {volume}
  {6}},\ \bibinfo {pages} {7503} (\bibinfo {year} {2015})},\ \Eprint
  {http://arxiv.org/abs/1411.6410} {arXiv:1411.6410 [gr-qc]} \BibitemShut
  {NoStop}%
\bibitem [{\citenamefont {Marin}\ \emph {et~al.}(2013)\citenamefont {Marin}
  \emph {et~al.}}]{Marin:2013pga}%
  \BibitemOpen
  \bibfield  {author} {\bibinfo {author} {\bibfnamefont {F.}~\bibnamefont
  {Marin}} \emph {et~al.},\ }\href {\doibase 10.1038/nphys2503} {\bibfield
  {journal} {\bibinfo  {journal} {Nature Phys.}\ }\textbf {\bibinfo {volume}
  {9}},\ \bibinfo {pages} {71} (\bibinfo {year} {2013})}\BibitemShut {NoStop}%
\bibitem [{\citenamefont {Marin}\ \emph {et~al.}(2014)\citenamefont {Marin},
  \citenamefont {Marino}, \citenamefont {Bonaldi}, \citenamefont {Cerdonio},
  \citenamefont {Conti}, \citenamefont {Falferi}, \citenamefont {Mezzena},
  \citenamefont {Ortolan}, \citenamefont {Prodi}, \citenamefont {Taffarello},
  \citenamefont {Vedovato}, \citenamefont {Vinante},\ and\ \citenamefont
  {Zendri}}]{Marin_2014}%
  \BibitemOpen
  \bibfield  {author} {\bibinfo {author} {\bibfnamefont {F.}~\bibnamefont
  {Marin}}, \bibinfo {author} {\bibfnamefont {F.}~\bibnamefont {Marino}},
  \bibinfo {author} {\bibfnamefont {M.}~\bibnamefont {Bonaldi}}, \bibinfo
  {author} {\bibfnamefont {M.}~\bibnamefont {Cerdonio}}, \bibinfo {author}
  {\bibfnamefont {L.}~\bibnamefont {Conti}}, \bibinfo {author} {\bibfnamefont
  {P.}~\bibnamefont {Falferi}}, \bibinfo {author} {\bibfnamefont
  {R.}~\bibnamefont {Mezzena}}, \bibinfo {author} {\bibfnamefont
  {A.}~\bibnamefont {Ortolan}}, \bibinfo {author} {\bibfnamefont {G.~A.}\
  \bibnamefont {Prodi}}, \bibinfo {author} {\bibfnamefont {L.}~\bibnamefont
  {Taffarello}}, \bibinfo {author} {\bibfnamefont {G.}~\bibnamefont
  {Vedovato}}, \bibinfo {author} {\bibfnamefont {A.}~\bibnamefont {Vinante}}, \
  and\ \bibinfo {author} {\bibfnamefont {J.-P.}\ \bibnamefont {Zendri}},\
  }\href {\doibase 10.1088/1367-2630/16/8/085012} {\bibfield  {journal}
  {\bibinfo  {journal} {New Journal of Physics}\ }\textbf {\bibinfo {volume}
  {16}},\ \bibinfo {pages} {085012} (\bibinfo {year} {2014})}\BibitemShut
  {NoStop}%
\bibitem [{\citenamefont {Westphal}\ \emph {et~al.}(2021)\citenamefont
  {Westphal}, \citenamefont {Pfaff},\ and\ \citenamefont {Aspelmeyer}}]{WHP21}%
  \BibitemOpen
  \bibfield  {author} {\bibinfo {author} {\bibfnamefont {H.}~\bibnamefont
  {Westphal}, \bibfnamefont {T.~andHepach}}, \bibinfo {author} {\bibfnamefont
  {J.}~\bibnamefont {Pfaff}}, \ and\ \bibinfo {author} {\bibfnamefont
  {M.}~\bibnamefont {Aspelmeyer}},\ }\href {\doibase
  10.1038/s41586-021-03250-7} {\bibfield  {journal} {\bibinfo  {journal}
  {Nature}\ }\textbf {\bibinfo {volume} {591}},\ \bibinfo {pages} {225}
  (\bibinfo {year} {2021})},\ \Eprint {http://arxiv.org/abs/2009.09546 [gr-qc]}
  {arXiv:2009.09546 [gr-qc]} \BibitemShut {NoStop}%
\bibitem [{\citenamefont {Bekenstein}(2012)}]{Bekenstein:2012yy}%
  \BibitemOpen
  \bibfield  {author} {\bibinfo {author} {\bibfnamefont {J.~D.}\ \bibnamefont
  {Bekenstein}},\ }\href {\doibase 10.1103/PhysRevD.86.124040} {\bibfield
  {journal} {\bibinfo  {journal} {Phys. Rev. D}\ }\textbf {\bibinfo {volume}
  {86}},\ \bibinfo {pages} {124040} (\bibinfo {year} {2012})},\ \Eprint
  {http://arxiv.org/abs/1211.3816} {arXiv:1211.3816 [gr-qc]} \BibitemShut
  {NoStop}%
\bibitem [{\citenamefont {Bekenstein}(2014)}]{Bekenstein:2013ih}%
  \BibitemOpen
  \bibfield  {author} {\bibinfo {author} {\bibfnamefont {J.~D.}\ \bibnamefont
  {Bekenstein}},\ }\href {\doibase 10.1007/s10701-014-9779-z} {\bibfield
  {journal} {\bibinfo  {journal} {Found. Phys.}\ }\textbf {\bibinfo {volume}
  {44}},\ \bibinfo {pages} {452} (\bibinfo {year} {2014})},\ \Eprint
  {http://arxiv.org/abs/1301.4322} {arXiv:1301.4322 [gr-qc]} \BibitemShut
  {NoStop}%
\bibitem [{\citenamefont {Petruzziello}\ and\ \citenamefont
  {Illuminati}(2021)}]{Petruzziello:2020wkd}%
  \BibitemOpen
  \bibfield  {author} {\bibinfo {author} {\bibfnamefont {L.}~\bibnamefont
  {Petruzziello}}\ and\ \bibinfo {author} {\bibfnamefont {F.}~\bibnamefont
  {Illuminati}},\ }\href {\doibase 10.1038/s41467-021-24711-7} {\bibfield
  {journal} {\bibinfo  {journal} {Nature Commun.}\ }\textbf {\bibinfo {volume}
  {12}},\ \bibinfo {pages} {4449} (\bibinfo {year} {2021})},\ \Eprint
  {http://arxiv.org/abs/2011.01255} {arXiv:2011.01255 [gr-qc]} \BibitemShut
  {NoStop}%
\bibitem [{\citenamefont {Moradpour}\ \emph {et~al.}(2019)\citenamefont
  {Moradpour}, \citenamefont {Ziaie}, \citenamefont {Ghaffari},\ and\
  \citenamefont {Feleppa}}]{Moradpour:2019wpj}%
  \BibitemOpen
  \bibfield  {author} {\bibinfo {author} {\bibfnamefont {H.}~\bibnamefont
  {Moradpour}}, \bibinfo {author} {\bibfnamefont {A.~H.}\ \bibnamefont
  {Ziaie}}, \bibinfo {author} {\bibfnamefont {S.}~\bibnamefont {Ghaffari}}, \
  and\ \bibinfo {author} {\bibfnamefont {F.}~\bibnamefont {Feleppa}},\ }\href
  {\doibase 10.1093/mnrasl/slz098} {\bibfield  {journal} {\bibinfo  {journal}
  {Mon. Not. Roy. Astron. Soc.}\ }\textbf {\bibinfo {volume} {488}},\ \bibinfo
  {pages} {L69} (\bibinfo {year} {2019})},\ \Eprint
  {http://arxiv.org/abs/1907.12940} {arXiv:1907.12940 [gr-qc]} \BibitemShut
  {NoStop}%
\bibitem [{\citenamefont {Easther}\ \emph {et~al.}(2001)\citenamefont
  {Easther}, \citenamefont {Greene}, \citenamefont {Kinney},\ and\
  \citenamefont {Shiu}}]{Easther:2001fi}%
  \BibitemOpen
  \bibfield  {author} {\bibinfo {author} {\bibfnamefont {R.}~\bibnamefont
  {Easther}}, \bibinfo {author} {\bibfnamefont {B.~R.}\ \bibnamefont {Greene}},
  \bibinfo {author} {\bibfnamefont {W.~H.}\ \bibnamefont {Kinney}}, \ and\
  \bibinfo {author} {\bibfnamefont {G.}~\bibnamefont {Shiu}},\ }\href {\doibase
  10.1103/PhysRevD.64.103502} {\bibfield  {journal} {\bibinfo  {journal} {Phys.
  Rev. D}\ }\textbf {\bibinfo {volume} {64}},\ \bibinfo {pages} {103502}
  (\bibinfo {year} {2001})},\ \Eprint {http://arxiv.org/abs/hep-th/0104102}
  {arXiv:hep-th/0104102} \BibitemShut {NoStop}%
\bibitem [{\citenamefont {Mureika}(2019)}]{Mureika_2019}%
  \BibitemOpen
  \bibfield  {author} {\bibinfo {author} {\bibfnamefont {J.}~\bibnamefont
  {Mureika}},\ }\href {\doibase 10.1016/j.physletb.2018.12.009} {\bibfield
  {journal} {\bibinfo  {journal} {Phys. Lett. B}\ }\textbf {\bibinfo {volume}
  {789}},\ \bibinfo {pages} {88} (\bibinfo {year} {2019})}\BibitemShut
  {NoStop}%
\bibitem [{\citenamefont {{Atkinson}}(1936)}]{atkinson1936relation}%
  \BibitemOpen
  \bibfield  {author} {\bibinfo {author} {\bibfnamefont {E.~C.}\ \bibnamefont
  {{Atkinson}}}\ }(\bibinfo {year} {1936})\ pp.\ \bibinfo {pages}
  {606--617}\BibitemShut {NoStop}%
\bibitem [{\citenamefont {Quesne}\ and\ \citenamefont
  {Tkachuk}(2010)}]{Quesne_2010}%
  \BibitemOpen
  \bibfield  {author} {\bibinfo {author} {\bibfnamefont {C.}~\bibnamefont
  {Quesne}}\ and\ \bibinfo {author} {\bibfnamefont {V.~M.}\ \bibnamefont
  {Tkachuk}},\ }\href {\doibase 10.1103/physreva.81.012106} {\bibfield
  {journal} {\bibinfo  {journal} {Phys. Rev. A}\ }\textbf {\bibinfo {volume}
  {81}} (\bibinfo {year} {2010}),\ 10.1103/physreva.81.012106}\BibitemShut
  {NoStop}%
\bibitem [{\citenamefont {Anderson}\ \emph {et~al.}(2023)\citenamefont
  {Anderson} \emph {et~al.}}]{anderson}%
  \BibitemOpen
  \bibfield  {author} {\bibinfo {author} {\bibfnamefont {E.~K.}\ \bibnamefont
  {Anderson}} \emph {et~al.} (\bibinfo {collaboration} {ALPHA}),\ }\href
  {\doibase 10.1038/s41586-023-06527-1} {\bibfield  {journal} {\bibinfo
  {journal} {Nature}\ }\textbf {\bibinfo {volume} {621}},\ \bibinfo {pages}
  {716} (\bibinfo {year} {2023})}\BibitemShut {NoStop}%
\bibitem [{\citenamefont {Gross}\ and\ \citenamefont {Wilczek}(1973)}]{Gross}%
  \BibitemOpen
  \bibfield  {author} {\bibinfo {author} {\bibfnamefont {D.}~\bibnamefont
  {Gross}}\ and\ \bibinfo {author} {\bibfnamefont {F.}~\bibnamefont
  {Wilczek}},\ }\href {\doibase 10.1103/PhyRevLett.30.1346G} {\bibfield
  {journal} {\bibinfo  {journal} {Phys. Rev. Lett.}\ }\textbf {\bibinfo
  {volume} {30}},\ \bibinfo {pages} {1343} (\bibinfo {year}
  {1973})}\BibitemShut {NoStop}%
\bibitem [{\citenamefont {Sivaram}\ and\ \citenamefont
  {Sinha}(1977)}]{Sivaram:1975dt}%
  \BibitemOpen
  \bibfield  {author} {\bibinfo {author} {\bibfnamefont {C.}~\bibnamefont
  {Sivaram}}\ and\ \bibinfo {author} {\bibfnamefont {K.~P.}\ \bibnamefont
  {Sinha}},\ }\href {\doibase 10.1103/PhysRevD.16.1975} {\bibfield  {journal}
  {\bibinfo  {journal} {Phys. Rev. D}\ }\textbf {\bibinfo {volume} {16}},\
  \bibinfo {pages} {1975} (\bibinfo {year} {1977})}\BibitemShut {NoStop}%
\bibitem [{\citenamefont {Di\'osi}(1984)}]{Diosi:1984wuz}%
  \BibitemOpen
  \bibfield  {author} {\bibinfo {author} {\bibfnamefont {L.}~\bibnamefont
  {Di\'osi}},\ }\href {\doibase 10.1016/0375-9601(84)90397-9} {\bibfield
  {journal} {\bibinfo  {journal} {Phys. Lett. A}\ }\textbf {\bibinfo {volume}
  {105}},\ \bibinfo {pages} {199} (\bibinfo {year} {1984})},\ \Eprint
  {http://arxiv.org/abs/1412.0201} {arXiv:1412.0201 [quant-ph]} \BibitemShut
  {NoStop}%
\bibitem [{\citenamefont {Penrose}(2014)}]{Penrose2014-PENOTG-2}%
  \BibitemOpen
  \bibfield  {author} {\bibinfo {author} {\bibfnamefont {R.}~\bibnamefont
  {Penrose}},\ }\href {\doibase 10.1007/s10701-013-9770-0} {\bibfield
  {journal} {\bibinfo  {journal} {Foundations of Physics}\ }\textbf {\bibinfo
  {volume} {44}},\ \bibinfo {pages} {557} (\bibinfo {year} {2014})}\BibitemShut
  {NoStop}%
\bibitem [{\citenamefont {Jusufi}\ and\ \citenamefont
  {Ali}(2023)}]{Jusufi:2023tdn}%
  \BibitemOpen
  \bibfield  {author} {\bibinfo {author} {\bibfnamefont {K.}~\bibnamefont
  {Jusufi}}\ and\ \bibinfo {author} {\bibfnamefont {A.~F.}\ \bibnamefont
  {Ali}},\ }\href@noop {} {\enquote {\bibinfo {title} {{Generalized Uncertainty
  Principle from the Regularized Self-Energy}},}\ } (\bibinfo {year} {2023}),\
  \bibinfo {note} {{}},\ \Eprint {http://arxiv.org/abs/2303.07198}
  {arXiv:2303.07198 [hep-th]} \BibitemShut {NoStop}%
\bibitem [{\citenamefont {Mureika}\ and\ \citenamefont
  {Mann}(1996{\natexlab{a}})}]{Mureika:1995ap}%
  \BibitemOpen
  \bibfield  {author} {\bibinfo {author} {\bibfnamefont {J.~R.}\ \bibnamefont
  {Mureika}}\ and\ \bibinfo {author} {\bibfnamefont {R.~B.}\ \bibnamefont
  {Mann}},\ }\href {\doibase 10.1016/0370-2693(95)01489-6} {\bibfield
  {journal} {\bibinfo  {journal} {Phys. Lett. B}\ }\textbf {\bibinfo {volume}
  {368}},\ \bibinfo {pages} {112} (\bibinfo {year} {1996}{\natexlab{a}})},\
  \Eprint {http://arxiv.org/abs/hep-ph/9511220} {arXiv:hep-ph/9511220}
  \BibitemShut {NoStop}%
\bibitem [{\citenamefont {Mureika}\ and\ \citenamefont
  {Mann}(1996{\natexlab{b}})}]{Mureika:1996de}%
  \BibitemOpen
  \bibfield  {author} {\bibinfo {author} {\bibfnamefont {J.~R.}\ \bibnamefont
  {Mureika}}\ and\ \bibinfo {author} {\bibfnamefont {R.~B.}\ \bibnamefont
  {Mann}},\ }\href {\doibase 10.1103/PhysRevD.54.2761} {\bibfield  {journal}
  {\bibinfo  {journal} {Phys. Rev. D}\ }\textbf {\bibinfo {volume} {54}},\
  \bibinfo {pages} {2761} (\bibinfo {year} {1996}{\natexlab{b}})},\ \Eprint
  {http://arxiv.org/abs/hep-ph/9603335} {arXiv:hep-ph/9603335} \BibitemShut
  {NoStop}%
\bibitem [{\citenamefont {Mureika}(1997)}]{Mureika:1996ud}%
  \BibitemOpen
  \bibfield  {author} {\bibinfo {author} {\bibfnamefont {J.~R.}\ \bibnamefont
  {Mureika}},\ }\href {\doibase 10.1103/PhysRevD.56.2408} {\bibfield  {journal}
  {\bibinfo  {journal} {Phys. Rev. D}\ }\textbf {\bibinfo {volume} {56}},\
  \bibinfo {pages} {2408} (\bibinfo {year} {1997})},\ \Eprint
  {http://arxiv.org/abs/hep-ph/9612391} {arXiv:hep-ph/9612391} \BibitemShut
  {NoStop}%
\bibitem [{\citenamefont {Ali}\ \emph {et~al.}(2022)\citenamefont {Ali},
  \citenamefont {Elmashad},\ and\ \citenamefont {Mureika}}]{Ali:2022ckm}%
  \BibitemOpen
  \bibfield  {author} {\bibinfo {author} {\bibfnamefont {A.~F.}\ \bibnamefont
  {Ali}}, \bibinfo {author} {\bibfnamefont {I.}~\bibnamefont {Elmashad}}, \
  and\ \bibinfo {author} {\bibfnamefont {J.}~\bibnamefont {Mureika}},\ }\href
  {\doibase 10.1016/j.physletb.2022.137182} {\bibfield  {journal} {\bibinfo
  {journal} {Phys. Lett. B}\ }\textbf {\bibinfo {volume} {831}},\ \bibinfo
  {pages} {137182} (\bibinfo {year} {2022})},\ \Eprint
  {http://arxiv.org/abs/2205.14009} {arXiv:2205.14009 [physics.gen-ph]}
  \BibitemShut {NoStop}%
\bibitem [{\citenamefont {Ali}\ \emph {et~al.}(2024)\citenamefont {Ali},
  \citenamefont {Elmashad}, \citenamefont {Mureika},\ and\ \citenamefont
  {Vagenas}}]{elmashad}%
  \BibitemOpen
  \bibfield  {author} {\bibinfo {author} {\bibfnamefont {A.}~\bibnamefont
  {Ali}}, \bibinfo {author} {\bibfnamefont {I.}~\bibnamefont {Elmashad}},
  \bibinfo {author} {\bibfnamefont {J.}~\bibnamefont {Mureika}}, \ and\
  \bibinfo {author} {\bibfnamefont {E.}~\bibnamefont {Vagenas}},\ }\href@noop
  {} {\enquote {\bibinfo {title} {{Theoretical and Observational Implications
  of Planck's Constant as a Running Fine Structure Constant}},}\ } (\bibinfo
  {year} {2024}),\ \bibinfo {note} {{Manucript in Preparation}}\BibitemShut
  {NoStop}%
\bibitem [{\citenamefont {{Ohanian}}(1986)}]{1986AmJPh..54..500O}%
  \BibitemOpen
  \bibfield  {author} {\bibinfo {author} {\bibfnamefont {H.~C.}\ \bibnamefont
  {{Ohanian}}},\ }\href {\doibase 10.1119/1.14580} {\bibfield  {journal}
  {\bibinfo  {journal} {American Journal of Physics}\ }\textbf {\bibinfo
  {volume} {54}},\ \bibinfo {pages} {500} (\bibinfo {year} {1986})}\BibitemShut
  {NoStop}%
\bibitem [{\citenamefont {{Belinfante}}(1939)}]{1939Phy.....6..887B}%
  \BibitemOpen
  \bibfield  {author} {\bibinfo {author} {\bibfnamefont {F.~J.}\ \bibnamefont
  {{Belinfante}}},\ }\href {\doibase 10.1016/S0031-8914(39)90090-X} {\bibfield
  {journal} {\bibinfo  {journal} {Physica}\ }\textbf {\bibinfo {volume} {6}},\
  \bibinfo {pages} {887} (\bibinfo {year} {1939})}\BibitemShut {NoStop}%
\bibitem [{\citenamefont {Christodoulou}\ and\ \citenamefont
  {Ruffini}(1971)}]{Ruffini}%
  \BibitemOpen
  \bibfield  {author} {\bibinfo {author} {\bibfnamefont {D.}~\bibnamefont
  {Christodoulou}}\ and\ \bibinfo {author} {\bibfnamefont {R.}~\bibnamefont
  {Ruffini}},\ }\href {\doibase 10.1103/PhysRevD.4.3552} {\bibfield  {journal}
  {\bibinfo  {journal} {Phys. Rev. D.}\ }\textbf {\bibinfo {volume} {4}},\
  \bibinfo {pages} {3552} (\bibinfo {year} {1971})}\BibitemShut {NoStop}%
\bibitem [{\citenamefont {Martinez}(1994)}]{Martinez:1994ja}%
  \BibitemOpen
  \bibfield  {author} {\bibinfo {author} {\bibfnamefont {E.~A.}\ \bibnamefont
  {Martinez}},\ }\href {\doibase 10.1103/PhysRevD.50.4920} {\bibfield
  {journal} {\bibinfo  {journal} {Phys. Rev. D}\ }\textbf {\bibinfo {volume}
  {50}},\ \bibinfo {pages} {4920} (\bibinfo {year} {1994})},\ \Eprint
  {http://arxiv.org/abs/gr-qc/9405033} {arXiv:gr-qc/9405033} \BibitemShut
  {NoStop}%
\bibitem [{\citenamefont {{Morris}}\ \emph {et~al.}(1988)\citenamefont
  {{Morris}}, \citenamefont {{Thorne}},\ and\ \citenamefont
  {{Yurtsever}}}]{1988PhRvL..61.1446M}%
  \BibitemOpen
  \bibfield  {author} {\bibinfo {author} {\bibfnamefont {M.~S.}\ \bibnamefont
  {{Morris}}}, \bibinfo {author} {\bibfnamefont {K.~S.}\ \bibnamefont
  {{Thorne}}}, \ and\ \bibinfo {author} {\bibfnamefont {U.}~\bibnamefont
  {{Yurtsever}}},\ }\href {\doibase 10.1103/PhysRevLett.61.1446} {\bibfield
  {journal} {\bibinfo  {journal} {Phys. Rev. Lett.}\ }\textbf {\bibinfo
  {volume} {61}},\ \bibinfo {pages} {1446} (\bibinfo {year}
  {1988})}\BibitemShut {NoStop}%
\bibitem [{\citenamefont {Casadio}\ and\ \citenamefont
  {Scardigli}(2014)}]{Cas13}%
  \BibitemOpen
  \bibfield  {author} {\bibinfo {author} {\bibfnamefont {R.}~\bibnamefont
  {Casadio}}\ and\ \bibinfo {author} {\bibfnamefont {F.}~\bibnamefont
  {Scardigli}},\ }\href {\doibase 10.1140/epjc/s10052-013-2685-2} {\bibfield
  {journal} {\bibinfo  {journal} {Eur. Phys. J. C}\ }\textbf {\bibinfo {volume}
  {74}},\ \bibinfo {pages} {2685} (\bibinfo {year} {2014})},\ \Eprint
  {http://arxiv.org/abs/1306.5298} {arXiv:1306.5298 [gr-qc]} \BibitemShut
  {NoStop}%
\bibitem [{\citenamefont {Knipfer}\ \emph {et~al.}(2019)\citenamefont
  {Knipfer}, \citenamefont {K\"oppel}, \citenamefont {Mureika},\ and\
  \citenamefont {Nicolini}}]{Knipfer:2019pgi}%
  \BibitemOpen
  \bibfield  {author} {\bibinfo {author} {\bibfnamefont {M.}~\bibnamefont
  {Knipfer}}, \bibinfo {author} {\bibfnamefont {S.}~\bibnamefont {K\"oppel}},
  \bibinfo {author} {\bibfnamefont {J.}~\bibnamefont {Mureika}}, \ and\
  \bibinfo {author} {\bibfnamefont {P.}~\bibnamefont {Nicolini}},\ }\href
  {\doibase 10.1088/1475-7516/2019/08/008} {\bibfield  {journal} {\bibinfo
  {journal} {JCAP}\ }\textbf {\bibinfo {volume} {08}},\ \bibinfo {pages} {008}
  (\bibinfo {year} {2019})},\ \Eprint {http://arxiv.org/abs/1905.03233}
  {arXiv:1905.03233 [gr-qc]} \BibitemShut {NoStop}%
\bibitem [{\citenamefont {{Carr}}(2013)}]{2013MPLA...2840011C}%
  \BibitemOpen
  \bibfield  {author} {\bibinfo {author} {\bibfnamefont {B.~J.}\ \bibnamefont
  {{Carr}}},\ }\href {\doibase 10.1142/S0217732313400117} {\bibfield  {journal}
  {\bibinfo  {journal} {Modern Physics Letters A}\ }\textbf {\bibinfo {volume}
  {28}},\ \bibinfo {eid} {1340011} (\bibinfo {year} {2013})}\BibitemShut
  {NoStop}%
\bibitem [{\citenamefont {{Witten}}(1981)}]{1981NuPhB.186..412W}%
  \BibitemOpen
  \bibfield  {author} {\bibinfo {author} {\bibfnamefont {E.}~\bibnamefont
  {{Witten}}},\ }\href {\doibase 10.1016/0550-3213(81)90021-3} {\bibfield
  {journal} {\bibinfo  {journal} {Nuclear Physics B}\ }\textbf {\bibinfo
  {volume} {186}},\ \bibinfo {pages} {412} (\bibinfo {year}
  {1981})}\BibitemShut {NoStop}%
\bibitem [{\citenamefont {{Witten}}(1995)}]{1995NuPhB.443...85W}%
  \BibitemOpen
  \bibfield  {author} {\bibinfo {author} {\bibfnamefont {E.}~\bibnamefont
  {{Witten}}},\ }\href {\doibase 10.1016/0550-3213(95)00158-O} {\bibfield
  {journal} {\bibinfo  {journal} {Nuclear Physics B}\ }\textbf {\bibinfo
  {volume} {443}},\ \bibinfo {pages} {85} (\bibinfo {year} {1995})},\ \Eprint
  {http://arxiv.org/abs/hep-th/9503124} {arXiv:hep-th/9503124 [hep-th]}
  \BibitemShut {NoStop}%
\bibitem [{\citenamefont {Nicolini}(ming)}]{aurilia}%
  \BibitemOpen
  \bibfield  {author} {\bibinfo {author} {\bibfnamefont {P.}~\bibnamefont
  {Nicolini}},\ }\enquote {\bibinfo {title} {{From Hadronic Bubbles to Quantum
  Gravity: Chasing the Infinite with Antonio Aurilia}},}\ in\ \href@noop {}
  {\emph {\bibinfo {booktitle} {{Touring the Planck Scale: Antonio Aurilia
  Memorial Volume}}}},\ \bibinfo {series and number} {Fundamental Theories of
  Physics},\ \bibinfo {editor} {edited by\ \bibinfo {editor} {\bibfnamefont
  {P.}~\bibnamefont {Nicolini}}}\ (\bibinfo  {publisher} {Springer Nature},\
  \bibinfo {year} {forthcoming})\ Chap.~\bibinfo {chapter} {{}}\BibitemShut
  {NoStop}%
\bibitem [{\citenamefont {{Mann}}(1985)}]{1985JMP....26.2308M}%
  \BibitemOpen
  \bibfield  {author} {\bibinfo {author} {\bibfnamefont {R.~B.}\ \bibnamefont
  {{Mann}}},\ }\href {\doibase 10.1063/1.526814} {\bibfield  {journal}
  {\bibinfo  {journal} {Journal of Mathematical Physics}\ }\textbf {\bibinfo
  {volume} {26}},\ \bibinfo {pages} {2308} (\bibinfo {year}
  {1985})}\BibitemShut {NoStop}%
\bibitem [{\citenamefont {{Gibbons}}\ and\ \citenamefont
  {{Wiltshire}}(1986)}]{1986AnPhy.167..201G}%
  \BibitemOpen
  \bibfield  {author} {\bibinfo {author} {\bibfnamefont {G.~W.}\ \bibnamefont
  {{Gibbons}}}\ and\ \bibinfo {author} {\bibfnamefont {D.~L.}\ \bibnamefont
  {{Wiltshire}}},\ }\href {\doibase 10.1016/S0003-4916(86)80012-4} {\bibfield
  {journal} {\bibinfo  {journal} {Annals of Physics}\ }\textbf {\bibinfo
  {volume} {167}},\ \bibinfo {pages} {201} (\bibinfo {year}
  {1986})}\BibitemShut {NoStop}%
\bibitem [{\citenamefont {{Cveti{\v{c}}}}\ and\ \citenamefont
  {{Youm}}(1995)}]{1995PhRvL..75.4165C}%
  \BibitemOpen
  \bibfield  {author} {\bibinfo {author} {\bibfnamefont {M.}~\bibnamefont
  {{Cveti{\v{c}}}}}\ and\ \bibinfo {author} {\bibfnamefont {D.}~\bibnamefont
  {{Youm}}},\ }\href {\doibase 10.1103/PhysRevLett.75.4165} {\bibfield
  {journal} {\bibinfo  {journal} {Phys Rev Lett}\ }\textbf {\bibinfo {volume}
  {75}},\ \bibinfo {pages} {4165} (\bibinfo {year} {1995})},\ \Eprint
  {http://arxiv.org/abs/hep-th/9503082} {arXiv:hep-th/9503082 [hep-th]}
  \BibitemShut {NoStop}%
\bibitem [{\citenamefont {Briet}\ and\ \citenamefont
  {Hobill}(2008)}]{Briet:2008mz}%
  \BibitemOpen
  \bibfield  {author} {\bibinfo {author} {\bibfnamefont {J.}~\bibnamefont
  {Briet}}\ and\ \bibinfo {author} {\bibfnamefont {D.}~\bibnamefont {Hobill}},\
  }\href@noop {} {\enquote {\bibinfo {title} {{Determining the Dimensionality
  of Spacetime by Gravitational Lensing}},}\ } (\bibinfo {year} {2008}),\
  \Eprint {http://arxiv.org/abs/0801.3859} {arXiv:0801.3859 [gr-qc]}
  \BibitemShut {NoStop}%
\end{thebibliography}%
%expects file "myrefs.bib"

\end{document}